**FOREWORD, JANUARY 2022**

The research documented in this manuscript was carried out in 1989, when the author was a principal investigator at Bell Laboratories. Although a seminal exposition of neutral atom beam focusing, the work did not strike a responsive chord with either the Physical Review journals or the Bell Labs management. After presentation in a contributed talk at the 1990 American Physical Society March Meeting, Mar 12-16, 1990, Anaheim, CA, and in an invited talk at the 1990 Annual Meeting of the Optical Society of America, Nov 4-9, 1990, Boston, MA, the manuscript was therefore laid aside to await further action. Action, as it turned out, that was three decades in the coming. In January 2022, Prof. Bodil Holst, Univ. of Bergen, Dept. of Physics and Technology, asked to cite the work in an upcoming review of atom microscopy. The original file was no longer to be found, but luckily preprints had been circulated in 1989. One had been sent by the author to Prof. J. Peter Toennies, director of the Max-Planck-Institut für Strömungsforschung in Göttingen, who had later forwarded a hardcopy to Prof Holst. She kindly scanned her copy of the preprint into a PDF and emailed it to the author. In this modern era the obvious course of action is to submit such a document to arXiv. Unfortunately arXiv expressly forbids *scanned* PDF files even under extraordinary circumstances. Accordingly the text of the scanned PDF was captured digitally, pasted into Microsoft Word, edited to remove transcription errors, and saved as this more enlightened form of PDF for uploading to arXiv. For completeness the *scanned* PDF of the original 1989 preprint has also been uploaded to Zenodo: Doak, Robert Bruce (2022). Focusing of an Atomic Beam (Version 1989). Zendo. (https://doi.org/.10.5821/zenodo.5855135).

# FOCUSSING OF AN ATOMIC BEAM

*R. B. Doak*

*AT&T Bell Laboratories*
*Murray Hill, New Jersey 07974*

*ABSTRACT*

A bent-crystal mirror has been used to focus an atomic helium beam. The mirror is made from gold deposited onto a mica substrate to form a thin film (~5000 Å thick) of large single crystal domains (domain size ~4000 Å ). The mica sheet is then bent *in situ* to form a cylindrical mirror of variable radius of curvature. Measurement of scattered beam intensity and angular distribution as a function of curvature demonstrate excellent focussing to within the mosaic spread of the surface. The reflectivity of the room temperature mirror is about ten percent. Potential uses are discussed, ranging from immediate applications in intensity and resolution enhancement of helium scattering experiments, to such long term possibilities as hyper-cooling of the beam or fabrication of a scanning atom microscope.

## INTRODUCTION

There are many reasons for scattering a neutral atom beam from a bent crystal surface. These may be divided into two general categories, depending upon whether the crystal bending is primarily intended to influence (1) the crystal properties (and in particular, those of the surface) or (2) the properties of the scattered beam. Underlying the first category is the recognition that bending produces a well-characterized stress distribution within the crystal which varies in sign and magnitude proportionally to the applied bending moments. Webb and co-workers[1] have demonstrated with electron diffraction that the static structure of a crystal surface can be modified through a surface stress applied in this fashion. Similar measurements could be made with atom diffraction. Moreover, with *inelastic* atom scattering[2] it should be possible to investigate the effects of surface stress on surface *vibrations* (surface phonons). This would allow a direct evaluation of the role of surface stress in determining surface vibrational properties. *In toto*, then, this first category comprises the "surface physics" experiments of atom scattering from bent crystals.

The second category is that of the "beam focussing" experiments. Here, as in optical focussing, the objective is to magnify or de-magnify the beam for the purpose of changing image size or beam spot size. Some simple applications of single mirror atomic beam focussing are sketched schematically in Fig. 1. Note that geometry (c) could be used as a rudimentary atom microscope with diffraction contrast by rastering a focussed beam spot across a target surface and monitoring the specular reflection or a diffraction peak. Analogues of imaging microscopes are also possible but less tractable in the absence of any atom equivalent to the phosphor screens or channel-plates used for large area detection of electrons and light.

Bending the target itself (in the absence of a mirror) will produce similar focussing effects. In general, the bending effects and focussing effects are inseparable. To investigate the surface physics it is necessary to take the bending-induced focusing properly into account; to construct an atom microscope from bent-crystal mirrors it is imperative that mirror surface and the atom scattering therefrom be well-characterized. The present manuscript describes the first experimental attempts to address these issues. The intent is (1) to lay the ground work for subsequent helium scattering measurements of surface structure and phonons in bent, stressed surfaces and (2) to quantify insofar as possible the feasibility of constructing a scanning atom microscope.

**EXPERIMENTAL APPARATUS**

These measurements were made with a high resolution helium atomic beam machine designed for studies of surface phonons. The bent-crystal reflector and its bending mechanism occupy the sample manipulator which would otherwise hold the target surface for phonon measurements. This arrangement and the pertinent dimensions are shown in Fig. 2. The apparatus[3] and the helium beam[4,5] have been described in some detail elsewhere. Most of these measurements were made with a liquid-nitrogen-cooled beam source, at which temperature the incident beam velocity was 945 m/sec and the velocity spread $\Delta v_{\parallel}/\bar{v} = 6 \times 10^{-3}$, FWHM. The corresponding De Broglie wavelength was $\lambda_{He} = 1.05 \text{Å}^{-1}$ or equivalent to that of a 136 eV electron.

The 0.81 mm diameter of the incident beam collimator subtends 0.29° with respect to the source. The actual beam full angle will be somewhat larger due to the finite size of the effective beam source (see below). This collimation produces a spot diameter of about 3 mm at the mirror surface and about 9 mm at the detector plane (for specular reflection). It is this beam spread upon which the focussing capabilities of the mirror are tested. The present measurements were limited to one-dimensional focussing using a cylindrical mirror. As the mirror radius of curvature is varied, the reflected beam spot should distort to an elliptical shape as indicated in Fig. 2. This decreases the transverse spot width in the horizontal plane and increases the axial intensity at least until the spot width decreases to a value comparable to 3 mm diameter of the detector aperture (0.09° full angle with respect to the source). For measurements in this latter regime (and to test convolution effects in general) a 50 μm diameter aperture (0.002°, full angle) could be inserted into the beam at a point 119 mm upstream of the detector plane.

The transverse spot width of the reflected beam is measured by rotating the mirror about its vertical axis, as shown in Fig. 2, under a constant total scattering angle of 71.5°. The effect is to sweep the reflected specular spot across the detector aperture, the axis of the specular beam rotating through twice the angle of rotation $\theta_m$ of the target manipulator. For these very narrow beam spreads, the angular width of the specular peak in such an angular distribution will reflect directly the transverse spot width at the detector. This angular width (FWHM) and the peak intensity of the specular reflection were taken as the primary indicators to characterize focussing. Such angular distributions will be given below as a function of the manipulator polar angle $\theta_m$: the normal angle of incidence $\theta_I$ of the beam onto the mirror decreases with increasing $\theta_m$ at 1:1; the specular angle for a flat mirror was $\theta_m = 155.4°$.

**CRYSTAL BENDER**

The miniature crystal-bending device is shown in Fig. 3. Pertinent dimensions are given in the outline drawing of Fig. 4. This mechanism mounts directly onto our commercial UHV sample manipulator,[6] replacing the existing target plate. The target itself is supported within two slotted tantalum rods which can be counter-rotated via a differentially-threaded drive screw to apply equal bending moments at the ends of the target; this bends the crystal into a cylindrical shape of uniform radius of curvature.[7] The drive screw is actuated by a "UHV-screwdriver" mounted on a rotary vacuum feed through. This bender design relies upon slippage of the target within the slotted rods to relieve any neutral plane tensile stress which develops as the slotted rods are rotated. Such lateral stress turns out to be of little consequence for focusing applications but may have to be taken into consideration for phonon measurements (e.g. by measurements at both positive and negative radius of curvature to remove its contribution, which is always tensile.)

The target could be heated from the rear by electron bombardment. Thermocouples were attached to the bender frame but not directly to the mirror surface itself. Accurate LEED and Auger spectroscopy measurements were not possible as the bender drive mechanism precluded proper positioning of either of these devices. Helium scattering was therefore used as the primary surface diagnostic probe; for the purpose of atom focussing this is clearly the most desirable characterization technique in any event.

## ATOM MIRROR

*Any* solid surface will reflect thermal energy inert gas atoms provided its temperature is not so low as to promote condensation (and even surface temperature may not be a constraint in the case of quantum reflections[8]). The key to forming an atom mirror is to carefully tailor both the microscopic and the macroscopic structure of the surface to make the scattering as coherent and elastic as possible. Enumerable light atom scattering experiments[9-11] have demonstrated that an appreciable fraction (> 50%) of an atomic beam incident upon a solid surface can be scattered in a coherent, elastic fashion provided: (1) The surface is clean and well-ordered on an atomic scale. (2) The ratio of beam atom mass to surface atom mass is small and the zero-point motion of the surface atoms is negligible. (3) The fraction of beam energy due to the component of momentum perpendicular to the surface is small relative to the surface Debye temperature $\theta_D{}^s$. (4) The surface temperature is low with respect to $\theta_D{}^s$.

To avoid diffuse (incoherent) elastic scattering, and to exclude extraneous broadening of the scattered beam, it is important that: (5) the mosaic spread of the surface be negligible and (6) the size of the coherently-scattering domains on the surface be large relative to the de Broglie wavelength of the atoms. To provide as much signal intensity as possible an appreciable fraction of the elastically-scattered beam should emerge in a single or, at most, a very few diffraction channels. This requires a surface with (7[a]) very little corrugation, such that specular scattering dominates or (7[b]) a small real-space lattice (to limit the number of diffraction peaks) *plus* an atomic corrugation profile which places the classical rainbow maxima at the diffraction angles. Finally, for technical reasons, it is useful to work with a relatively inert surface such that once a clean, well-ordered surface has been prepared it will remain so for an extended period of time. There is considerable latitude within these constraints, i.e., an increase in beam temperature to adjust the atom de Broglie wavelength can be compensated for by a decrease in surface temperature. Indeed, these requirements are nothing more than an extended set of the conventional constraints for atom scattering investigations of surfaces.

Meeting the above constraints will maximize the reflectivity of the reflector surface. To focus the scattered beam it is necessary to shape the reflector to the proper macroscopic form without altering these microscopic properties. A given beam atom impinging onto a flat surface will sample a microscopic area of certainly no more than (and probably *much* less than) its first Fresnel zone, given by radius

$$R_f \;=\; \left( \frac{x_{sm}\, x_{md}}{x_{sd}} \, \lambda_{He} \right)^{1/2}$$

or about 6 $\mu$m in the present instance (subscripts s, m, and d denote source, mirror, and detector). A cut and polished crystal surface will therefore not produce focussing, but rather flat planar scattering from the microscopic facets convoluted with any inter-facet interference patterns.

To produce focussing there are several alternative approaches: First, a planar focussing element could be used. It may actually now be technically feasible to fabricate, for atom focussing, a planar array of reflecting Fresnel zone patterns,[12] either by electron beam lithography or possibly even STM direct writing. This would be slow, difficult, and would probably have to be done *in situ* to preserve the surface order and cleanliness.

A second approach is to fabricate the mirror from an amorphous rather than crystalline material. After cutting, polishing and cleaning, the surface must be coated with an overlayer or thin film to smooth out the atomic scale irregularities. Such an atom mirror was recently demonstrated by Berkhout, et al.[8] for liquid helium films on fused quartz substrates. While that particular system is ill-suited for general focussing applications due to the light mass and large zero point motion of the helium layer, the general approach may be viable, depending upon the atomic structure of the overlayer as discussed below.

A third possibility, and that employed here, is to bend a single crystal reflector to produce a focussing shape. This is a time-honored technique in both X-ray[13] and neutron[14,15] scattering. A bent crystal can produce, at the minimum, a piecewise approximation to a macroscopic cylindrical or spherical surface. Depending on the atomic details of strain relief within the crystal, the individual planar domains may bend as well: This is of particular interest for certain purposes (e.g. Fresnel interference need no longer limit the lateral coherence) and may be crucial for very high resolution focussing. Since the individual beam atoms are mutually incoherent (each atom wave packet interferes only with itself), the coherence length[16] of the beam is no larger than that of the individual atoms. Thus an array of planar facets conforming to a macroscopic cylindrical or spherical form is sufficient for focussing down to a spot size of the order of the surface domain size, the atom coherence length, or apparatus transfer width, whichever is limiting.

The present mirrors were fabricated from mica sheets. Single crystal gold thin films were grown upon the mica substrates to tailor the surface properties in accordance with the requirements detailed above. Gold is an admirable choice for this apart from its relatively low surface Debye temperature; this is largely compensated by the beneficially heavy gold atomic mass relative to the helium beam mass. Chidsey, Trevor, and co-workers[17,18] have documented in detail the procedures by which the gold films are deposited onto the mica substrates. Crucial for the present experiment is the structure of the individual crystallites comprising the film. To obtain the largest, flattest domains it is necessary to ramp the substrate temperature from around 120° C to about 250° C during the deposition. The results reported here are for a 5000 Å thick film grown upon a 200 $\mu$m thick mica substrate. The single crystal domains are about 4000 Å across and may have 8-10 atomic steps distributed over this distance. The contribution of this finite domain size to the overall spread of the scattered beam, expressed as a rotation of the target $\Delta\theta_t$ should be 0.008°, FWHM, for the incident beam wavelength of 1.05 Å. This is much smaller than other limiting factors, as will be seen below. In particular, the mosaic spread of the mica substrates is reported to be about 0.1°.

After deposition the mica sheets were cut to shape for mounting in the crystal bender. It was found that edge delamination could be largely avoided if the sheets were pressed between glass microscope slides during cutting. Nonetheless, some delamination was always present at the edges and it was attempted to avoid these regions during subsequent measurements.

Once in UHV the gold film was cleaned by argon ion sputtering at 500 V and 0.9 $\mu$A/cm$^2$ for 30 min with the sample at room temperature. The gold film would anneal slowly even at room temperature, its helium specular intensity rising with a time constant of about 1/2 hour in agreement with STM diffusion measurements.[19] This produced a relatively poor surface, however, judged on the basis of elastic and inelastic helium scattering. A modest anneal of about 15 minutes at an estimated temperature of 400±50 C (thermocouple reading 280 C on the bender frame) produced helium specular intensity, time-of-flight (TOF) definition, and incoherent elastic background comparable to those reported elsewhere.[20,21] When at this annealing temperature, the equilibrium specular intensity was about 1/4 of its room temperature value. The sputtering and annealing cycle was eventually repeated several times with no noticeable deterioration of the film. Specular helium scattering from a freshly cleaned surface would drop in amplitude by no more than a few percent over the course of 24 hours.

## LASER BEAM FOCUSSING

Focussing of visible light was tested using a HeNe laser.  The laser beam was passed through .a slightly canted neutral density filter to produce an array of diverging beams, as shown in Fig. 5.  The angular spread from the first to the fourth spot was about 0.36° or slightly more than the helium beam full angle spread.  Distances from laser to mirror and from mirror to detection plane roughly approximated the corresponding helium beam distances.  The scattering angle, however, was only about 5° as opposed to 71.5° for the beam apparatus.  By using an array of spots it is possible to remove any mosaic spread, which will broaden the individual spots but not affect their angular separation.

The specularly-reflected spot pattern is shown in Fig. 5 as a function of the number of rotations of the crystal bender drive screw (0 turns being nominally flat, 15 turns being the most concave).  Reasonable focussing is attained.  For this paraxial geometry, optimum focussing occurs at about 11 turns of the drive screw. The corresponding radius of curvature is about 1 m.  These data are replotted in a more quantitative fashion in Fig. 6, taking the separation of the first and fourth spot as a measure of the array spread.  Clearly the crystal does not bend substantially until the drive screw has been rotated about 5 or 6 turns.  Neglecting distortion at 14 turns and above, the plot is reasonable linear, as would be expected.  The slope, however, is only about half of the anticipated value, given the dimensions of Fig. 4. This may be due to tensile strain as mentioned above, or.to compressive strain at the lines of contact between crystal and bending rods.  There appears to be no adverse effect on the focussing, indicating that the crystal is nonetheless distorting into a reasonably cylindrical shape. Erect and inverted images of Fig. 5 may be discerned by the intensity variation which distinguishes the individual spots. This will not be true of a helium beam spread, for which only the magnitude of the beam width can be measured, as simulated by neglecting the polarity information (inverted triangles in Fig. 6). The crystal could be bent, unbent, and rebent with no sign of hysteresis, provided the last motion of the drive screw was consistently made in the same direction.

## ATOM BEAM FOCUSSING

Angular distributions of the specular helium atom reflection are shown in Fig. 7 as the crystal is bent from an unstressed, nominally flat shape (0 turns of drive screw) into an increasingly concave shape. The geometry is that of Fig. 2. The atom scattering is not paraaxial (total scattering angle of 71.5°). However, even in this off-axis geometry there should be substantial focussing (Appendix A). Alignment, in particular out-of-plane tilt, was checked at each setting of drive screw prior to recording the angular distribution. The specular amplitude rises to a maximum at about 10 turns, concurrent with a minimum angular peak width. Peak width and amplitude for this series are plotted in Fig. 8. It is evident that the atomic beam is being focussed fairly effectively. Inserting the much smaller 50 $\mu$m detector aperture into the beam produces little additional decrease in specular width, even at the minimum with l0 turns as shown in Fig. 9. This identifies the limiting width as that of the beam and not of the detector aperture. As such it is of interest to replot the angular width as in Fig. 10, adding a minus sign to those widths at and beyond 10 turns (triangles). If this is repeated after simple subtraction of 0.12° from all points, the result is a curve which passes smoothly through the abscissa with no break in slope (Fig.10, circles). This 0.12° width is attributed to the mosaic spread of the surface, in close agreement with the anticipated value (above).

The integrated area under the specular peak also offers insight into the details of the crystal bending. If the picture of a simple elliptical distortion as in Fig. 2 is accurate, the integrated specular intensity will remain constant as the crystal is bent. From Fig. 11 (triangles) this is seen to be true at and beyond about 8 turns of the drive screw i.e. once the target begins to bend substantially. At lesser bends there are clearly some astigmatic shape changes taking place. Some indication of this is also seen in the shapes of the reflected laser spots, Fig. 5. This may in part be due to the edge delamination problems mentioned above. Also plotted in Fig. 11 is the specular angle $\theta_m(00)$. This was taken from the average of the two FWHM points; the increase beyond 10 turns is associated with the specular peak becoming somewhat asymmetric. Note, however, that the total variation of $\theta_m(00)$ is less than 0.3° over the entire range. This implies the bend is reasonably symmetric about the target center.

## MIRROR ATOM REFLECTIVITY

In the present beam machine, the total scattering angle may be rotated to 180° to send the incident beam straight into the detector,[4] which allows the mirror reflectivity to be measured directly. This was done for an unstressed (0-turns, nominally flat) room temperature crystal. The beam source was at liquid nitrogen temperature ($\lambda_{He}$ = 1.0 Å) as throughout the scattering measurements. Using the quadrupole mass spectrometer detector (as with all of the above measurements), the ratio of specular intensity to incident beam intensity was 6.2%.

The apparatus incorporates a pitot chamber for making an absolute measurement of the incident beam intensity. This can also be used to monitor any scattered beam which is sufficiently intense, as is the case with specular scattering from the gold film. The specular to incident beam intensity ratio measured with the pitot chamber was 10±2%. The disagreement is a not unreasonable indicator of the uncertainties of the calibration. Being much closer to the target, the pitot reading will be less affected by the extraneous angular dispersion in the specular beam due to the target mosaic spread, and should produce a higher ratio.

**TIME-OF-FLIGHT MEASUREMENTS**

A bent crystal can focus a beam in velocity space as well as real space.[15] Such effects are not anticipated for elastic scattering in the present instance for two reasons. First of all, k-space focussing requires an anistropic velocity distribution across the angular spread of the beam, for example when the beam has been generated by diffraction from a crystal monochromator. Secondly, we are hopefully dealing with pure specular scattering which should produce no angular dispersion of the reflected beam due to any incident beam velocity spread, isotropic or otherwise. Time-of-flight (TOF) analysis is used to energy-analyze the helium beam in the present apparatus.[5] TOF spectra are shown in Fig. 12 for the specular reflection with the crystal bender at 0 turns (nominally flat) and at 10 turns (optimum focussing). The focussing is seen to enhance the specular amplitude while leaving its spectral lineshape unchanged, in agreement with expectations.

For inelastic scattering from a bent crystal, the situation becomes somewhat more complicated. TOF spectra taken two degrees off-specular are shown in Fig. 13. (The scales of both abscissa and ordinate have been changed from those of Fig. 12.) The narrow peak at 1790 $\mu$sec arises through incoherent elastic scattering, presumably from surface imperfections: This peak is about $10^{-3}$ of the specular amplitude of Fig. 12. As the surface is bent to focus the beam, the amplitude of this incoherent elastic peak increases and its width remains unchanged, displaying essentially the same behavior as the specular elastic peak of Fig. 12.

The broader peak at shorter flight times is due to energy gain (anti-Stokes) inelastic scattering. This scattering angle was chosen to minimize the accessible range of phonon momentum and energy so that, in principle, interactions with single surface phonons should produce TOF signatures almost as narrow as those of the elastic peaks. The striking feature is that this inelastic peak is virtually unchanged in TOF lineshape when the crystal is bent, as is easily seen in the difference spectrum at the bottom of Fig. 13.

In fact, as in neutron scattering,[22] one would expect significant narrowing and enhancement of the inelastic peak. This may be seen by plotting the "scan curve" for each ray within the incident beam; a scan curve gives the locus of those values of energy and momentum exchange which are consistent with conservation relations. Superimposing scan curves on the phonon dispersion curves allows a simple graphical representation of inelastic scattering.[23] Since diverging rays within the beam correspond to slightly different angles of scattering from the mirror surface, and since these angles will change as the crystal is bent, so too will the scan curves. The envelope of scan curves at the specular angle are displayed in Fig. 14 for a flat surface and for a surface bent for optimum focussing. These scan curves are drawn for a beam full angle of $3°$ (i.e., expanded by about a factor of 10 over the actual value) in order to yield a discernible spread for the focussed beam. If the target is rotated by a small amount, the first order effect is to shift the scan curves parallel to themselves along the abscissa. In this off-specular geometry, the scan curves for different rays within the beam will no longer pass through a single point of intersection on this axis. However, the caustic will remain small for near-specular angles. Taking into account the non-zero size of the source and the detector as well as the finite velocity distribution of the beam will also broaden the envelope of off-specular scan curves somewhat. These are relatively minor contributions in the present instance and one would therefore still expect the focussing to produce an appreciable enhancement of any inelastic TOF features that are due to single surface phonons. The absence of such enhancement in Fig. 13 may most likely be attributed to the in-plane mosaic spread of the gold crystallites within the film. Polar angular distributions taken at different azimuthal angles, as in Fig. 15, show diffraction features which are very broad in azimuth, indicating that the distribution of azimuthal orientations is broad. In effect, then, the inelastic scattering is not sampling just a single phonon dispersion curve, as in Fig. 14, but rather a whole family of curves corresponding to the different azimuthal directions. It is useful to enumerate other factors which may diminish any inelastic enhancement due to focussing. First, any contribution from bulk phonons or multiple phonon scattering will not be enhanced unless it involves an $\omega(\vec{Q})$ singularity (e.g., a resonance peak in the bulk bands or a sharp maximum in the scattering cross-section) which is narrower than the envelope of unfocussed scan curves. Second, the bending itself may distort the phonon dispersion curves in a non-uniform manner from one crystallite to the next, again producing a family of dispersion curves as with the mosaic spread above. Third, phonon lifetimes may limit the angular and energy width of the single surface phonon peaks. We remark that at typical phonon velocities of $10^3$ m/sec, broadening on the scale of I meV implies. phonon mean free paths of about 10 Å or less, so that the finite domain size should not be a problem in the present instance.

**SOURCE SIZE**

Up to this point, we have tacitly treated the helium nozzle as a point source situated at the nozzle exit. The fact that focussing can be demonstrated experimentally shows this to indeed be a reasonable assumption to the present degree of accuracy. For scanning atom microscope applications, however, the requirements are much more stringent. The effective size of the source (both in cross-sectional area and in depth of field) will ultimately fix the resolution of the microscope and it is useful to discuss this in some depth.

The effective source size is determined by the details of the supersonic free-jet expansion which generates the beam. The gas dynamics of supersonic free-jets have been discussed by a number of authors.[24-30] In a high Mach number jet (M>~2.5; compare terminal Mach numbers of $M_T$~280-330 in the present apparatus) the near-axial flow field more than a few nozzle diameters downstream of the nozzle exit closely resembles that of a simple source flow. The streamlines may be approximated as straight lines diverging radially outward from a virtual point source. For helium, the density at a point specified by (R,$\theta$) is[24]

$$\rho(R,\theta) \,=\, \rho(R,0)\, cos^2(1.15\theta) \tag{1}$$

where $\theta$ is the angle between streamline and axis. Along the streamlines the density and therefore the frequency of collisions drops rapidly due to the radial divergence. At some point each beam atom will undergo its last collision within the expansion, i.e. the beam enters free-molecular flow. If the width of this transition zone is sufficiently narrow, it may be modeled as occuring abruptly at a single spherical surface, the "sudden-freeze" surface,[25] with a radius $R_f$, for $M_T$>>5, of[28]

$$\frac{R_f}{D*} \cong \left(\frac{M_T}{A}\right)^{\frac{1}{\gamma-1}} \qquad (2)$$

Here $\gamma$ is the ratio of specific heats (1.63 for He), A a calculated constant (3.26 for He[24]), and D* the nozzle diameter. Atoms will leave a point on the sudden-freeze surface in directions narrowly distributed about the local normal and with velocities narrowly distributed above the median beam velocity

$$\bar{v} = \left(\frac{2\gamma kT}{(\gamma-1)m}\right)^{1/2} \frac{\vec{R}}{R} \qquad (3)$$

These distributions are specified by the perpendicular and parallel velocity distributions $f(v_\perp)$ and $f(v_{//})$, respectively, and often characterized in width by assigning temperatures $T_\perp$, $T_{//}$.[26,27] For atom microscopy applications, this sudden-freeze surface is the "emitting" surface: the "brightness" of different parts of the surface will be proportional to $\rho(R_f, \theta)$, eqn. (1), and the angular distribution of "emitted" atoms will be given by $f(v_\perp/\bar{v})$, assuming $v_\perp << \bar{v}$.

To establish the size and location of the effective virtual source, the free-molecular rays which describe the atom trajectories outside of the freeze-in surface are extended as virtual rays backwards in the direction of the nozzle.[26] The envelope of these rays will have a point of minimum cross-section somewhere between the continuum flow source point at $x_o$ and the "sudden freeze" surface at $x=R_f$. This cross-over acts as a virtual source for the downstream optics. Its diameter depends upon the angular spread $f(v_\perp/\bar{v})$ of rays leaving the freeze-in surface, on the radius $R_f$ of the freeze-in surface, and on the full-angle stop aperture $\alpha_i$ determined by the downstream optics. If $\alpha_i \gg \Delta v_\perp / \bar{v}$, the cross-over will be located at $x_o$ and have diameter $D_c \cong v_\perp R_f / \bar{v}$. If $\alpha_i \ll \Delta v_\perp / \bar{v}$, the cross-over will be at $x_f$ with $D_c \cong \alpha_i R_f$. Since the radial and parallel velocity components in the beam remain approximately equilibrated up to the sudden-freeze surface, we can estimate $\Delta v_\perp / \bar{v}$ from measured values of $\Delta v_{//} / \bar{v}$, about $5 \times 10^{-3}$ FWHM for a high-speed-ratio beam.[5] With a sudden-freeze radius of 5-15 mm, this corresponds to an effective source diameter of 25-75 $\mu$m, or comparable to corresponding cross-over diameters for electron beams generated by thermionic emission.[31] An actual measurement of the effective source size (Appendix B) gives a somewhat larger value of 330 $\mu$m. The discrepancy may partly be due to broadening by background laboratory vibrations. We have not yet made any attempt to accurately characterize or eliminate such vibrations, but would expect them to be much less than the ~200 $\mu$m length scales involved in the source size measurements. Thus, within the admittedly crude approximations of the modeling, it appears that $\Delta v_\perp \cong 5 \bar{v}$. Since all helium beam experiments to date have essentially ignored $\Delta v_\perp$ while minimizing $\Delta v_{//}$, this is perhaps to be expected. Measurements reported elsewheres[5,32] show that the effective source size increases with increasing stagnation pressure in the nozzle. This will partly be due to an increase in $R_f$ with Mach number but may include increases of $\Delta v_\perp$ as well. If so, $\Delta v_\perp$ could be reduced by operating at lower source stagnation pressures than commonly employed.

It is possible to decrease the effective source size by restricting the stop aperture, i.e. by setting $\alpha_i \ll \Delta v_\perp / \bar{v}$. This is quite detrimental to beam intensity, however. Reducing the effective diameter by a factor of 10 would lower the beam signal to around background levels in the present apparatus. It is much more effective to tailor the flow characteristics of the nozzle to produce as small an effective source as possible, then open the detector aperture to capture a large solid angle of the beam. There are potentially large steric factors involved here: As in the present apparatus, the detector typically subtends a few microsteradians with respect to the source. As a first approximation to the solid angle which might be captured (this will obviously require uniformity of flow characteristics across the entire angle) we take that solid angle subtended by the Mach disk[24,27] of diameter $D_m$ at position $x_m$. Experiments[33] have shown that

$$\frac{D_m}{x_m} \cong 0.5 \tag{4}$$

over a wide range of pressure ratios. The corresponding full angle aperture is about 30°, or a solid angle of about 0.2 sr, representing gain of about $10^5$ in intensity for the present apparatus.

We note in passing that the Mach disk location is given by[24]

$$\frac{x_m}{D*} = 0.67 \left(\frac{p_0}{p_1}\right)^{1/2} \tag{5}$$

where $p_0$ is the nozzle stagnation pressure and $p_1$ ambient pressure downstream of the nozzle. Typically our $p_0/p_1 > 10^6$, so $x_m > 670D*$. For a terminal Mach number of 300, Eqn. (2) yields $R_f \cong 130D*$ i.e. the beam achieves free molecular flow well before reaching the Mach disk, which is therefore a "virtual" Mach disk.[29]  In this situation it may be possible to capture an even larger solid angle of the beam.

The terminal Mach number for helium is expected to scale with nozzle diameter D* as[30]

$$M_T \sim (p_0 D^*)^{0.5 \sim 1} \tag{6}$$

Hence, from Eqns. (2) and (B3)

$$D_s = D^{*1 \sim 1.5} \tag{7}$$

The most effective way to reduce the effective source diameter is therefore to reduce the nozzle diameter.[34]  The trade-off comes in the need to decrease velocity resolution (which scales with $\sim M_T$) and beam intensity ($\sim p_0 D^{*2}$) to avoid an increase in clustering of the beam ($\sim p_0^2 D^*$).

## SOURCE INTENSITY AND BRIGHTNESS

The absolute beam intensity may be determined with a pitot chamber as described above.  Pitot pressure is measured with an ionization gauge which has been calibrated *in situ* against a rotating ball viscosity manometer.[35]  For beam conditions typical of the present focussing experiments (10 $\mu$m nominal nozzle diameter, liquid-nitrogen cooled, 85 bar) the absolute beam intensity was measured to be $7 \times 10^{19}$ atoms sr$^{-1}$ sec$^{-1}$.  With room temperature beams, intensities as high as $2 \times 10^{20}$ atoms sr$^{-1}$ sec$^{-1}$ have been obtained.  For comparison with electron sources,[31] these correspond to "currents" of 11 and 32 A sr$^{-1}$, respectively.  An analysis of the pitot aperture geometry shows the radial velocity distribution at the sudden-freeze surface to still be the limiting factor in determining the effective source size. Taking the value measured for the present beam (330 $\mu$m diameter) or that reported elsewhere for the room temperature beam[28] (180 μm) we calculate a source brightness of $8 \times 10^{22}$ and $7 \times 10^{23}$ atoms cm$^{-2}$ s$^{-1}$ sr$^{-1}$, respectively. In electron terms this would correspond to $1.3 \times 10^4$ and $1.1 \times 10^5$ A cm$^{-2}$ s$^{-1}$ sr$^{-1}$, respectively, or comparable to values obtained with thermionic electron emitters.[31]

## INTRA-BEAM SCATTERING AT IMAGE POINTS

Cross-sections for He+He scattering have been measured only down to relative velocities of about 70 m/see where, for $^4$He, the total cross-section is about 150 Å$^2$ and rising exponentially with decreasing velocity.[36] Such measurements are made by merging two beams at as small an enclosed angle as possible, typically about 20°. In essence, a focussed beam presents a very similar geometry, merging rays at angles ranging from 0° up to the full aperture angle of the beam. For the geometry of Fig. 2 the latter would be 0.16° for point source and detector (with perfect focussing), giving a *maximum* relative velocity of 2.7 m/sec. In principle then, with suitable masking to define two converging rays, it should be possible to utilize this geometry to measure H$^4$e+$^4$He total scattering cross-sections to far below the Ramsauer-Townsend minimum. In terms of helium microscopy it is conceivable that this converging geometry could introduce non-negligible effects, scattering atoms out of the beam and leading to attenuation and broadening of the focussed spot. Whether such intra-beam scattering is important will depend upon the density in the beam (i.e. upon the effective source size and brightness, the stop aperture of the optics, and the efficiency of focussing) as well as the relevant helium scattering cross-sections. In an extreme case it is perhaps not impossible that the beam could re-enter transition flow ("unfreeze") at an image point, conceivably even allowing additional cooling ("hyper-cooling") of the beam. Intra-beam scattering, should it prove to be a problem, could always be avoided by reducing the beam density, e.g. by lowering the nozzle pressure or by reducing the stop aperture of the beam. Effectively, then, this may fix an upper limit on usable beam intensity.

**MIRROR ATOMIC STRUCTURE**

There appears to be no alternative to the use of a solid surface focussing element for atom beam focussing. As with neutron optics[37] one can define a "refractive index" for atoms as

$$n(\vec{r}) = \left(1 - \frac{V(\vec{r})}{E}\right)^{1/2} \qquad (8)$$

where $V(\vec{r})$ is the potential energy function of the atom and E its energy. In the absence of an electric charge and dipole moment, only atomic scale interactions contribute to $V(\vec{r})$ for helium atoms. Macroscopic focussing fields, as with electric and magnetic lenses for charged particles, are simply not possible.

The complication which arises is that the figures of merit of the atom mirror must then encompass its atomic structure as well as its macroscopic form. Whether this becomes a limiting factor depends upon the application. For near-unity magnification or demagnification, as in the present measurements or the simple applications depicted in Fig. 1, this is not a major concern. For atom microscope applications, which would require demagnification of $10^3$ or more, the atomic structure of the mirror will be decisive. Since a "piecewise" approximation to a cylindrical or spherical surface can only focus down to a spot size comparable to the "piece" size, i.e. the domain size of the crystallites on the mirror surface, it would seem that sub-micron atom microscopy will require the atomic surface to exhibit the same curvature as the macroscopic form. Whether it is possible to attain this remains to be seen. As far as we are aware, this has never been investigated although it would seem to be within the realm of present STM or helium scattering measurements.

**CONCLUSION**

We have demonstrated the use of bent-crystal reflecting optics to focus an atomic beam. It appears that this could be immediately applied to enhance intensity and/or angular resolution of helium atom scattering as depicted in Fig.1. We anticipate no problems in removing the present constraint of mirror mosaic spread, either by working with substrates other than mica or by growing the gold thin films on cut and polished fused quartz substrates, nor with scaling up the present mirror to capture a larger solid angle of the beam. There are potentially large gains to be realized. Considering, for example, the intensity enhancement of Fig. 1a, it is possible with a spherical mirror to increase the solid angle of the beam by

$$\left(2 - \frac{x_m}{x_d}\right)^2 \left(\frac{x_d}{x_m}\right)^2 \tag{9}$$

without changing the angular resolution of the apparatus (the focused beam beam *converges* on the target surface at the same angle as the *divergence* of the unfocussed beam). For a reasonable source-to-mirror distance of 100 mm and using the dimensions of the present apparatus this yields a factor of 1300 increase. With minor sacrifices in angular resolution this could, in addition, be focussed onto target areas of <500 $\mu$m. The obvious extension would be to helium atom microscopy of solid surfaces, taking advantage of the strict surface sensitivity, the benign and inert character, and the high absolute energy resolution of helium atom scattering. The above measurements show that source size and brightness need not be limiting factors unless intra-beam scattering restricts the usable intensity of the beam. (One must bear in mind, however, that helium atom detection efficiencies are *much* lower than those of electrons.) Successful helium microscopy will depend critically on the atomic scale structure of the atom reflector. Data on the domain size and atomic-scale curvature of the bent crystal surfaces or that of thin films deposited onto cut and polished amorphous substrates are not yet available but should be within the realm of present experimental techniques.

**ACKNOWLEDGEMENTS**

We are indebted to C. E. Chidsey for growing the gold-on-mica sample used in this study, to D. J. Trevor for STM analysis of its surface topography, and to both for many useful discussions. We wish to thank E. E. Chaban for his skillful assistance in constructing the crystal bending mechanism.

APPENDIX A

We wish to characterize the degree of focussing in the non-paraxial geometry of Fig. 16. A beam from a point source at O scatters off of a cylindrical mirror of radius R. The beam source lies neither on the mirror axis nor on the Rowland circle. Consider a ray emerging at angle $\phi$ to the beam axis, where $\phi$ can be either positive (as shown) or negative. The focussing of the reflected beam is characterized by calculating w, the distance between the reflected ray and the reflected beam axis in the direction perpendicular to the latter. Given the definitions of Fig. 16 and considering specular scattering only ($\theta_f = \theta_i$, $\beta_f = \beta_i$) it is straightforward to show that

$$\frac{w}{\cos\theta_i} = y^*[1 - \tan\theta_i \tan(\phi - 2\gamma)]$$

$$- x^*[\ \tan\theta_i\ +\ \tan(\phi - 2\gamma)] \qquad (1)$$

$$+ x^*[1\ +\ (\tan\theta_i)^2 \tan(\phi - 2\gamma)]$$

Define

$$r = R\cos\theta_i \qquad (2)$$

and note that the Rowland circle is given by

$$\xi_0 = \xi = r \qquad (3)$$

For small values of $\phi$ and for $\theta_i$ not too large, it is easily shown that, to $O(\phi^2)$,

$$x^* \cong \frac{\xi^2}{2r\cos\theta_i}\phi^2 \tag{4}$$

$$y^* \cong \frac{\xi_0}{\cos\theta_i}\left[\phi + \tan\theta_i\left(1 - \frac{\xi_0}{2r}\right)\phi^2\right] \tag{5}$$

$$\phi - 2\gamma \cong \left(1 - \frac{2\xi_0}{r}\right)\phi - \frac{2\xi_0\tan\theta_i}{r}\left(1 - \frac{\xi_0}{2r}\right)\phi^2 \tag{6}$$

whence

$$w = \left(\xi_0 + \xi - \frac{2\xi\xi_0}{r}\right)\phi$$

$$+ \frac{\xi_0\tan\theta_i}{r}\left(\xi_0 - 2\xi + \frac{\xi\xi_0}{r}\right)\phi^2 \tag{7}$$

and the constraints on d and di are that

$$\phi \ll 1 \tag{8}$$

$$\theta_i \ll \frac{\pi}{2} - \frac{\xi_0}{R}$$

The image point I lies at that value of $\xi$ for which w vanishes to at least $O(\phi)$;

$$\frac{1}{\xi_I} + \frac{1}{\xi_0} = \frac{2}{r} \tag{9}$$

This is the familiar concave mirror formula with an effective radius of curvature r. If $\xi_0$ lies on the Rowland circle, then so must $\xi_I$ (as seen from Eqns. (3) and (9)) and moreover w vanishes to $O(\phi^2)$ at $\xi_I$. For general $\xi_0$, $\xi_I$, the reflected beam full width at $\xi = \xi_I$ is

$$w_I = \left[\frac{3\xi_I}{16}\left(\frac{\xi_0^2}{\xi_I^2} - 1\right)\tan\theta_i\right]\alpha^2 \tag{10}$$

where $\alpha$ is the full angular spread of the beam. Given the dimensions and scattering angle of Fig. 2, this yields $w_l = -3\ \mu m$ as opposed to 9 mm without focussing. There is thus substantial focussing even with a relatively large scattering angle and for points off of the Rowland circle.

To plot scan curves it is necessary to know $\beta_i,\ \beta_f$ as a function of $\phi$. For specular scattering ($\beta_i = \beta_f$) from the bent crystal this is, to second order in

$$\beta_i \cong \theta_i + \tfrac{1}{2}\left(1 - \frac{\xi_0}{\xi_I}\right)\phi - \frac{\tan\theta_i}{8}\left(1 + \frac{\xi_0}{\xi_I}\right)\left(3 - \frac{\xi_0}{\xi_I}\right)\phi^2 \qquad (11)$$

and the total scattering angle for each $\phi$ is simply

$$\theta_{sd}(\phi) = \beta_i(\phi) + \beta_f(\phi) = 2\beta_i(\phi) \qquad (12)$$

For comparison, the analogous relations for scattering from a flat surface are (for general $\theta_i,\ \theta_f$)

$$\beta_i = \theta_i + \phi$$

$$\beta_f = tan^{-1}\left[\tan\theta_f - \frac{\xi_0}{\xi_I}\left(\frac{\sin\phi}{\cos\theta_f\cos(\theta_i + \phi)}\right)\right] \qquad (13)$$

APPENDIX B

Since the helium beam is in free-molecular flow outside of the sudden-freeze surface, the effective source size is easily measured by aperturing[4,26] or blocking the beam in this region. The latter approach is demonstrated here. The geometry is illustrated schematically in Fig. 17. Dimensions are those of Fig. 2. The small detector aperture ($D_d = 50$ $\mu$m diameter) was used; it is located a distance $x_d = 1715$ mm from the nozzle. The sudden-freeze model assumes continuum radial source flow within the sudden-freeze radius $R_f$ followed by an abrupt transition to free-molecular flow. Atoms leaving the sudden-freeze surface will be narrowly distributed in angle about the local normal to the surface with an angular distribution given by $f(v_\perp/\bar{v})$ where $f(v_\perp)$ is the perpendicular velocity distribution. We characterize this angular spread with $\Delta v_\perp/\bar{v}$ where $\Delta v_\perp$ is the full width of the velocity distribution. The angles of importance are $\alpha_s$, the angle subtended by the effective source diameter $D_s$ with respect to a point in the detector plane, and $\alpha_d$, the angle subtended by the detector diameter $D_d$ with respect to the virtual source at $x = x_o \cong 0$. From the geometry of Fig. 17 it is easily seen that

$$\alpha_s = \frac{R_f}{x_d}\frac{\Delta v_\perp}{\bar{v}} \tag{1}$$

Note that the effective source diameter at $R_f$ as seen by a point detector would be

$$D_s = R_f\left(1 - \frac{R_f}{x_d}\right)\frac{\Delta v_\perp}{\bar{v}} \tag{2}$$

so that at most ($x_d \gg R_f$) an area on the sudden-freeze surface of extent

$$D_S \cong R_f \frac{\Delta v_\perp}{\bar{v}} \tag{3230}$$

will contribute. As the target is moved transversely into the beam, the detected intensity drops to zero, as plotted in Fig. 18. The derivative of this intensity curve gives the beam profile at the target plane. The measured width is $\Delta y_t = 230$ $\mu$m for a liquid-nitrogen-cooled source at 85 bar using a 10 $\mu$m nominal diameter nozzle. From the geometry of the extremal rays in Fig. 17

$$D_S = \frac{(x_d - R_f)\Delta y_t - (x_t - x_f)D_d}{(x_d - x_t)} \tag{4}$$

This yields an effective source size of $D_s = 330$ $\mu$m giving a transverse velocity width of $\Delta v_\perp / \bar{v} = 2\%$. This estimate ignores the details of properly convoluting a realistic angular distribution $f(\Delta v_\perp / \bar{v})$ with a planar target edge obscuring a circular detector orifice, but such details should not appreciably change the calculated value.

**FIGURE CAPTIONS**

Figure 1.   Possible applications of single mirror focussing in helium scattering experiments.   Shaded area depicts unfocussed beam (all atomic beam sources produce radially divergent flow fields).   (a) Signal enhancement by capturing a larger solid angle of beam (aperture stop becomes mirror or lens size rather than detector size).   (b) Time-of-flight resolution enhancement by chopping at intermediate crossover point to decrease chopper shutter function.   (c) Selected area sampling by focussing onto the target surface.

Figure 2   Schematic diagram of the experimental apparatus.   A tightly-collimated, high-speed-ratio, helium atomic beam is scattered from a cylindrical mirror of variable radius of curvature.   Only the limiting apertures are shown.   The specular beam cross-section should change from circular (solid lines) to elliptical (dashed lines) as the mirror is bent (dashed lines).   At optimum focus, an ideal mirror would image the horizontal extent of the effective source onto the detector plane with a magnification given by the ratio of the mirror-detector to source-mirror distances.   A 50 $\mu$m diameter aperture can be inserted. at a point 119 mm upstream of detector plane.

Figure 3.   Target bending mechanism.   Target crystal (long rectangular piece) passes through slots in pair of vertical bending rods.   Rods counter-rotate when drive screw at top is turned with UHV

"wobble stick" mounted on vacuum feedthrough. The bends the target into cylindrical form which can be either concave (CCW rotation of drive screw) or convex (CW).

Figure 4.   Outline drawing of drive mechanism with pertinent dimensions in inches. Drive screw is differentially threaded to provide very small effective pitch (448 threads/inch).

Figure 5.   Optical focussing characteristics of gold-on-mica thin film using HeNe laser spot array.  Laser-mirror and mirror-detector plane distances were 0.64 m and 2.69 m, respectively, with an enclosed angle of about 5°. Specular spot patterns are labeled by the number of turns CCW of drive screw on crystal bender.  Crystal is unstressed at 0 turns and becomes concave with ever decreasing radius of curvature as drive screw is rotated CCW.  Graph paper is 6 divisions/inch.

Figure 6.   Measured spread between first and fourth reflected laser spots, taken from the data of Fig. 5, as a function of drive screw rotation on crystal bender. Inverted triangles show results if sign information is neglected.

Figure 7.   Angular distribution of specular helium reflection as a function of drive screw rotation of crystal bender.  Mirror is being rotated while holding fixed the total angle of 71.5° between incident beam axis and target-detector axis.  Detector emission current 0.5 mA.

Figure 8.   Angular width and peak amplitude of specular helium reflection as crystal is bent, taken from the data of Fig. 7.  Optimum focusing occurs at about 10 turns of drive screw.  Replication error bars are shown for widths at 0 and 10 turns.

Figure 9.   With drive screw set at 10 turns, angular distribution of specular helium reflection was recorded both with the 3 mm detector aperture (solid lines) and 50 $\mu$m detector aperture (dashed lines).  Detector emission current 0.5 and 5 mA, respectively.  No additional narrowing occurs with smaller aperture.

Figure 10.  Angular width of helium specular reflection adjusted by:  (1) adding negative polarity to points at and beyond crossover point at 10 turns (triangles) and (2) subtracting 0.12° from all measured widths before switching sign at 10 turns and beyond (circles).

Figure 11.  In-plane integrated specular helium intensity (circles) and manipulator angle of specular peak (triangles) as a function of number of drive screw rotations.

Figure 12.  Time-of-flight spectra for specular reflection under optimum focussing (top; 10 turns of drive screw) and with a flat surface (middle; 0 turns).  Difference spectrum is shown at bottom.  TOF and detection parameters are given; chopper to detector distance was 1.692 m.

Figure 13.  Inelastic TOF spectra at two degrees off of the specular peak.

TOF and detection parameters, as shown, are different from those of Fig. 12. (The incoherent elastic peak at 1790 $\mu$sec is a factor of $10^3$ lower in amplitude than the specular TOF peak.) Crystal bender was again set to give optimum focussing (top; 10 turns) and a flat surface (middle; 0 turns). Bottom trace is difference spectrum: inelastic signature is essentially absent in difference trace.

Figure 14. Scan curve depiction of focussing in $\omega(\vec{Q})$ space. Envelopes of scan curves are drawn for all rays passing from point source to point detector for focussed and for unfocussed beam. Parameters are those of the experimental measurements except that the beam angular spread was taken to be 3°, or roughly 10 times the actual value. Data points and theoretical dispersion curves from Harten, et a1.[20] for Au(111) are shown in the irreducible Brillouin zone; these curves should be reflected across both axes and repeated periodically to give the full set of intersections of scan curves and dispersion curves.

Figure 15. Polar angular scans of scattered intensity as the sample is rotated in azimuth about its normal for a nominally unbent surface. The displacement of the diffraction feature from the nominal ($\bar{1}0$) angle is probably a result of the soliton reconstruction of the Au(111)surface.[21]

Figure 16.  Geometrical definitions for scattering from source O to image point I from a cylindrical mirror of radius R.  A particular ray at angle $\phi$ to the beam axis is traced.

Figure 17.  Geometrical definitions for discussion of finite size effects.  The transverse velocity distribution at the "sudden-freeze" surface $R_f$ is given by $f(v_\perp/\bar{v})$ and characterized here by a FWHM $\Delta v_\perp/\bar{v}$.

Figure 18.  Beam blocking to ascertain source size.  Measurements are for a Iiquid-nitrogen-cooled beam, 85 bar, 10 pm nom. diameter using the 50 $\mu$m diameter collimator at the detector.  With the target rotated to give a slightly negative angle of incidence, it is moved transversely into the incident beam to attenuate it.  Both dataset (triangles) and its pointwise derivative (dots) are shown.

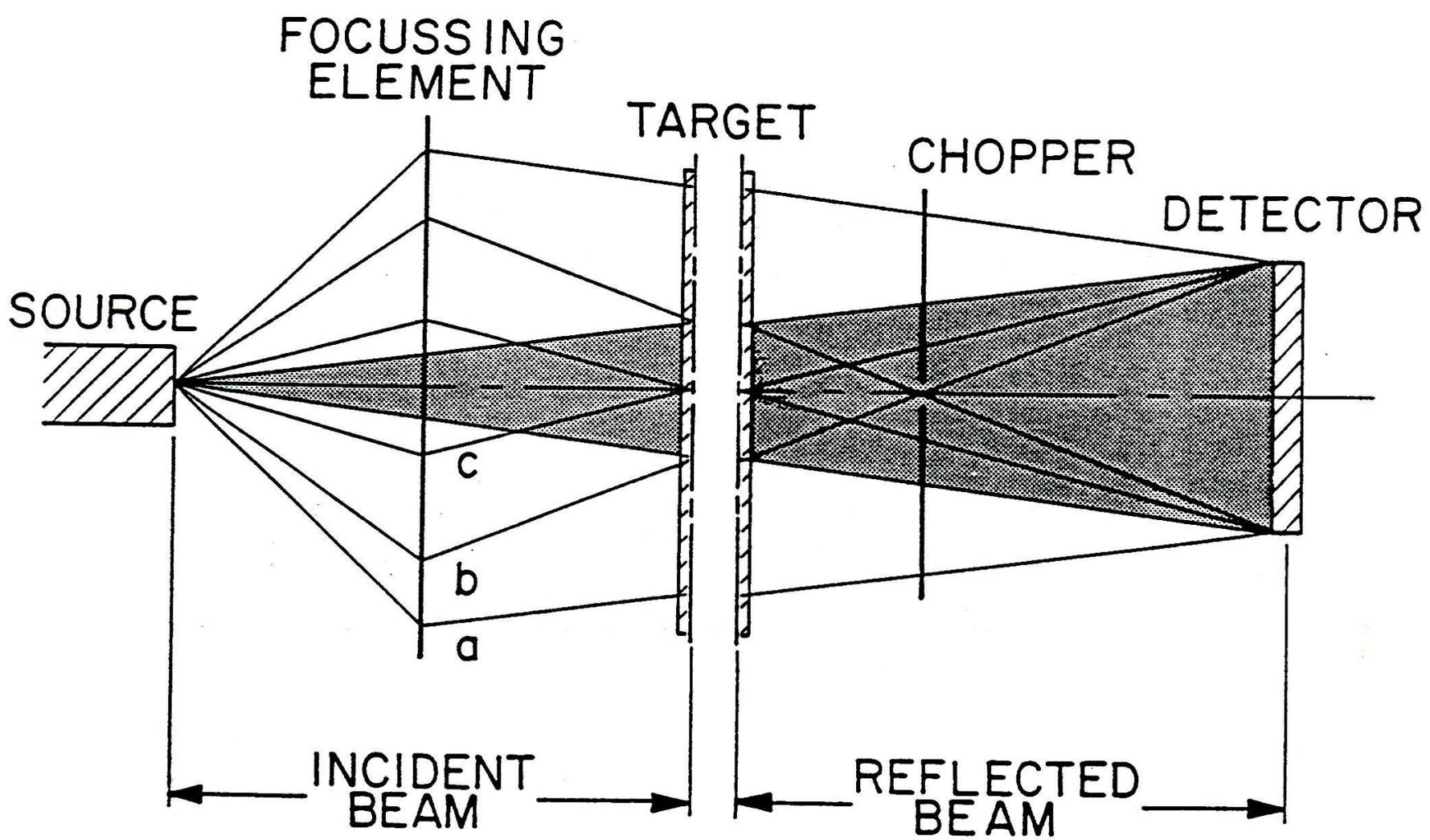

FOCUSSING
ELEMENT

TARGET

CHOPPER

DETECTOR

SOURCE

c

b

a

INCIDENT
BEAM

REFLECTED
BEAM

DOAK ①

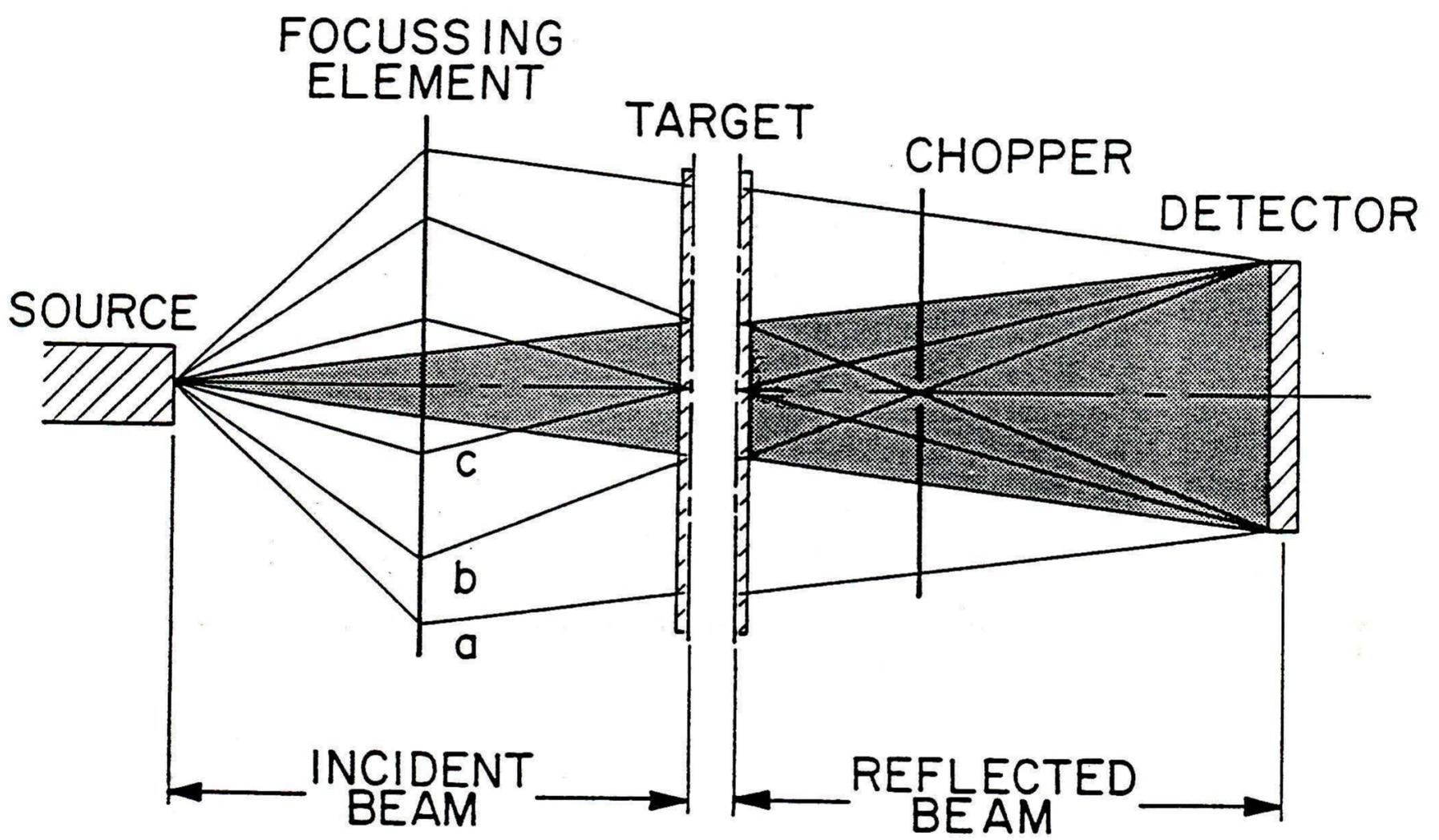

FOCUSSING
ELEMENT

TARGET

CHOPPER

DETECTOR

SOURCE

c

b

a

INCIDENT
BEAM

REFLECTED
BEAM

DOAK 1

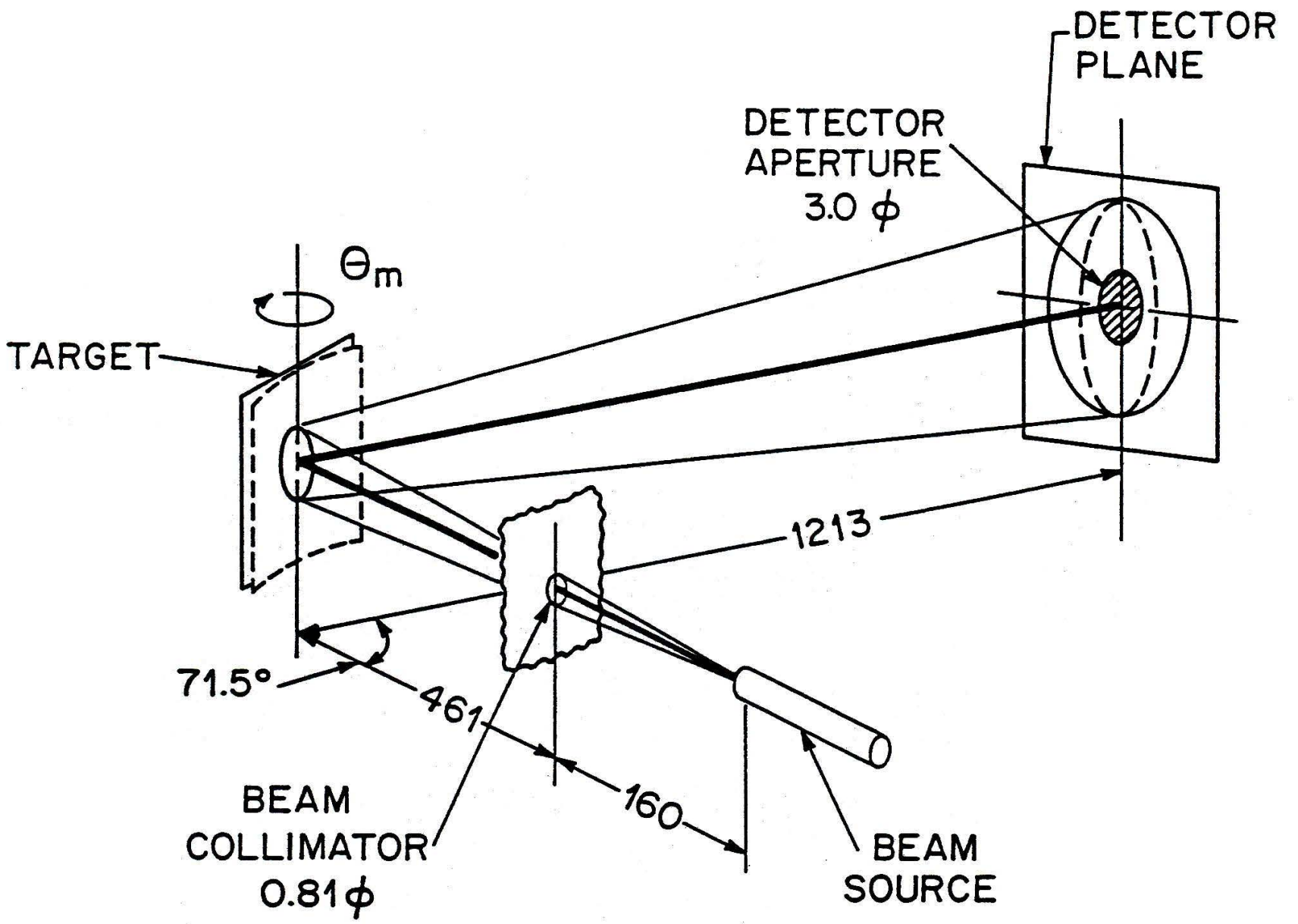

DETECTOR
PLANE

DETECTOR
APERTURE
3.0 ϕ

$\Theta_m$

TARGET

1213

71.5°

461

160

BEAM
COLLIMATOR
0.81ϕ

BEAM
SOURCE

ALL DIMENSIONS mm

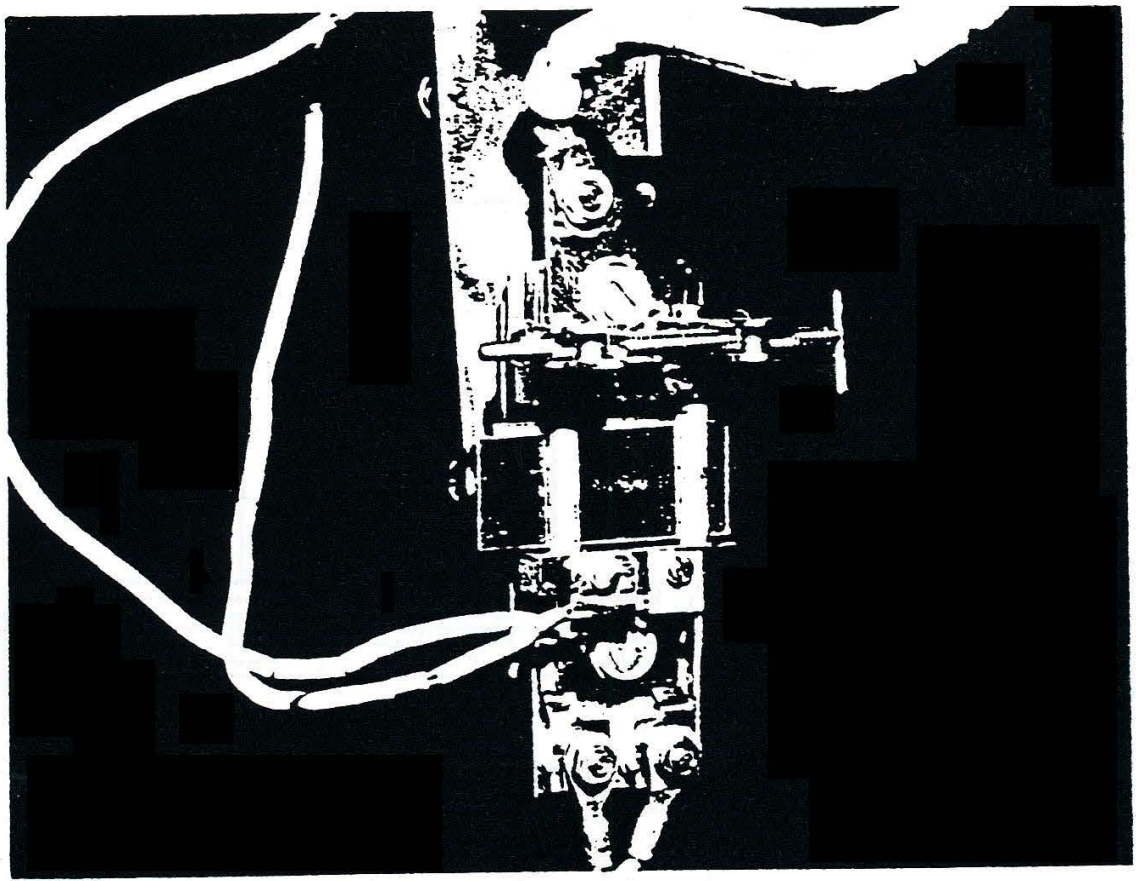

DOAK ③

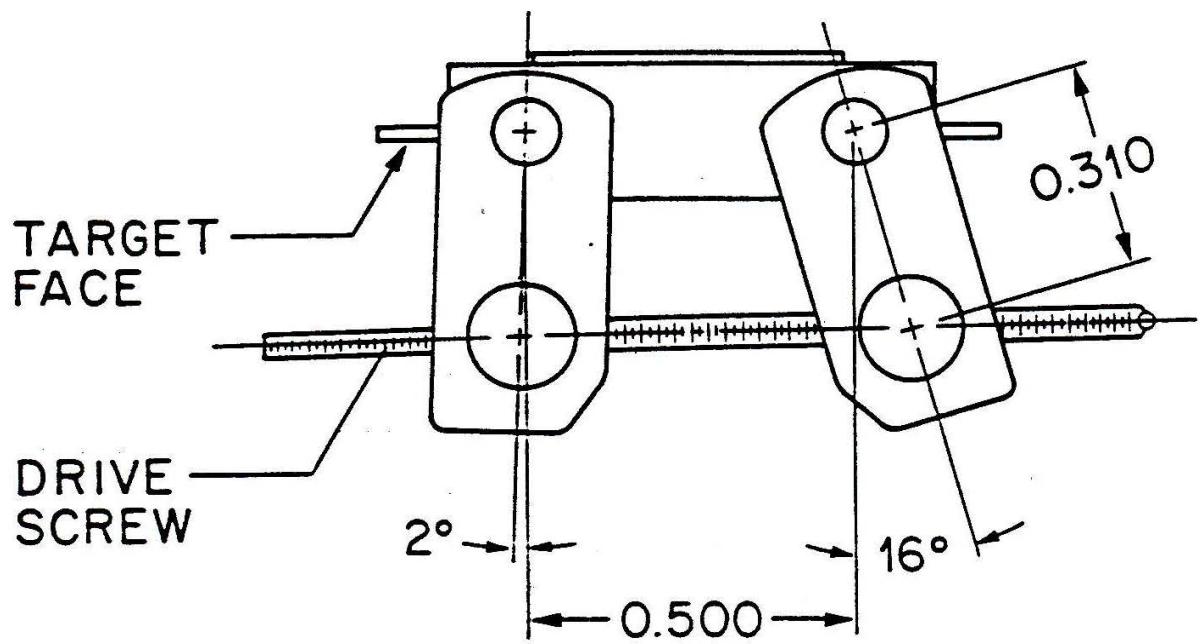

TARGET
FACE

DRIVE
SCREW

0.310

2°  16°

0.500

1/4 INCH

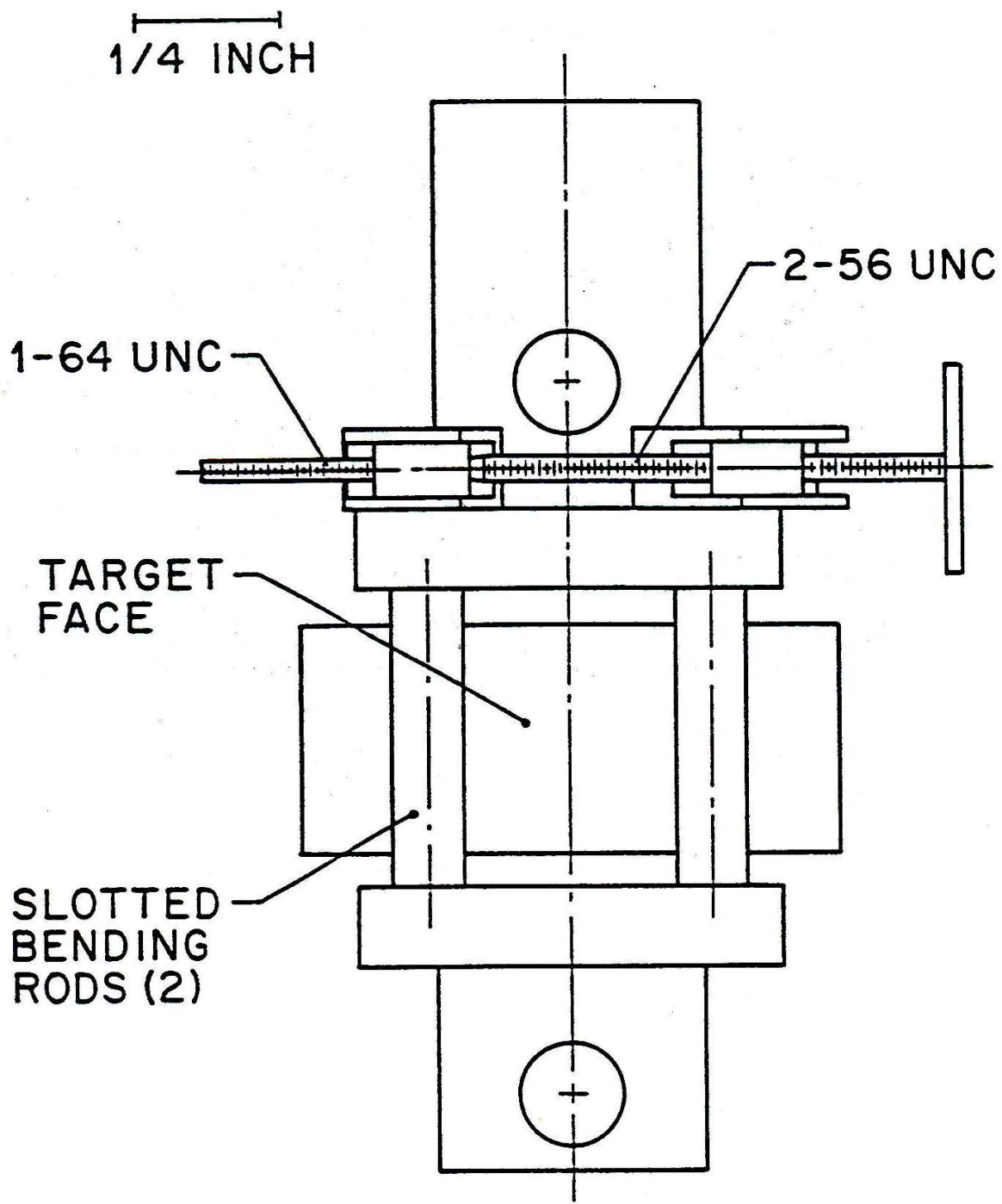

2-56 UNC

1-64 UNC

TARGET
FACE

SLOTTED
BENDING
RODS (2)

DOAK ④

INCIDENT BEAM

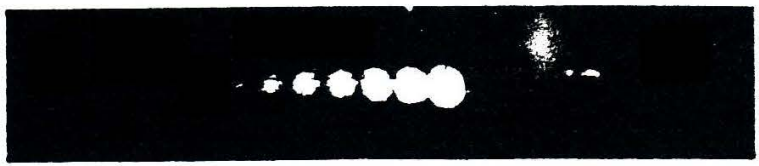

REFLECTED BEAM

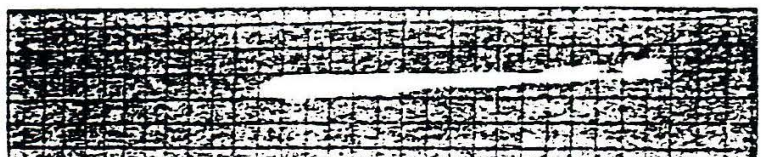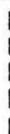

0 TURNS

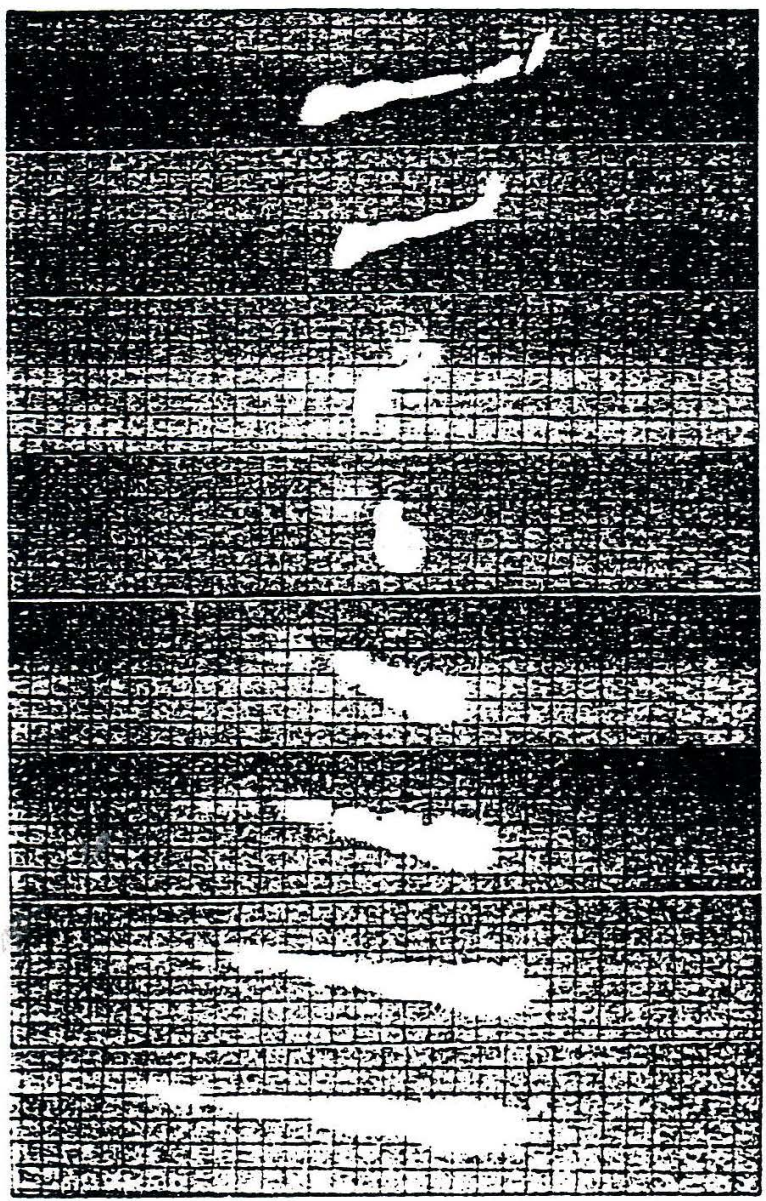

DOAK (5)

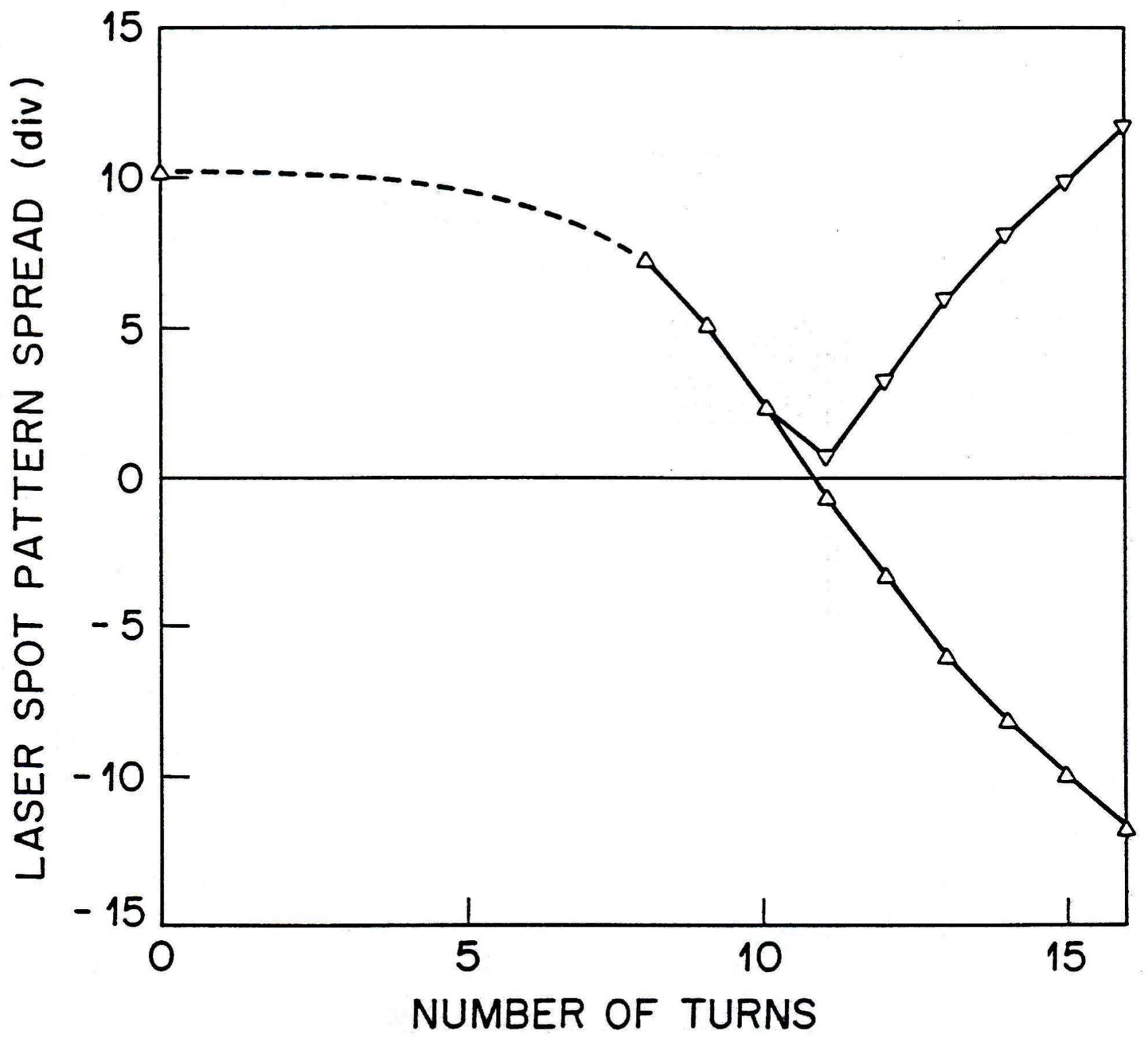

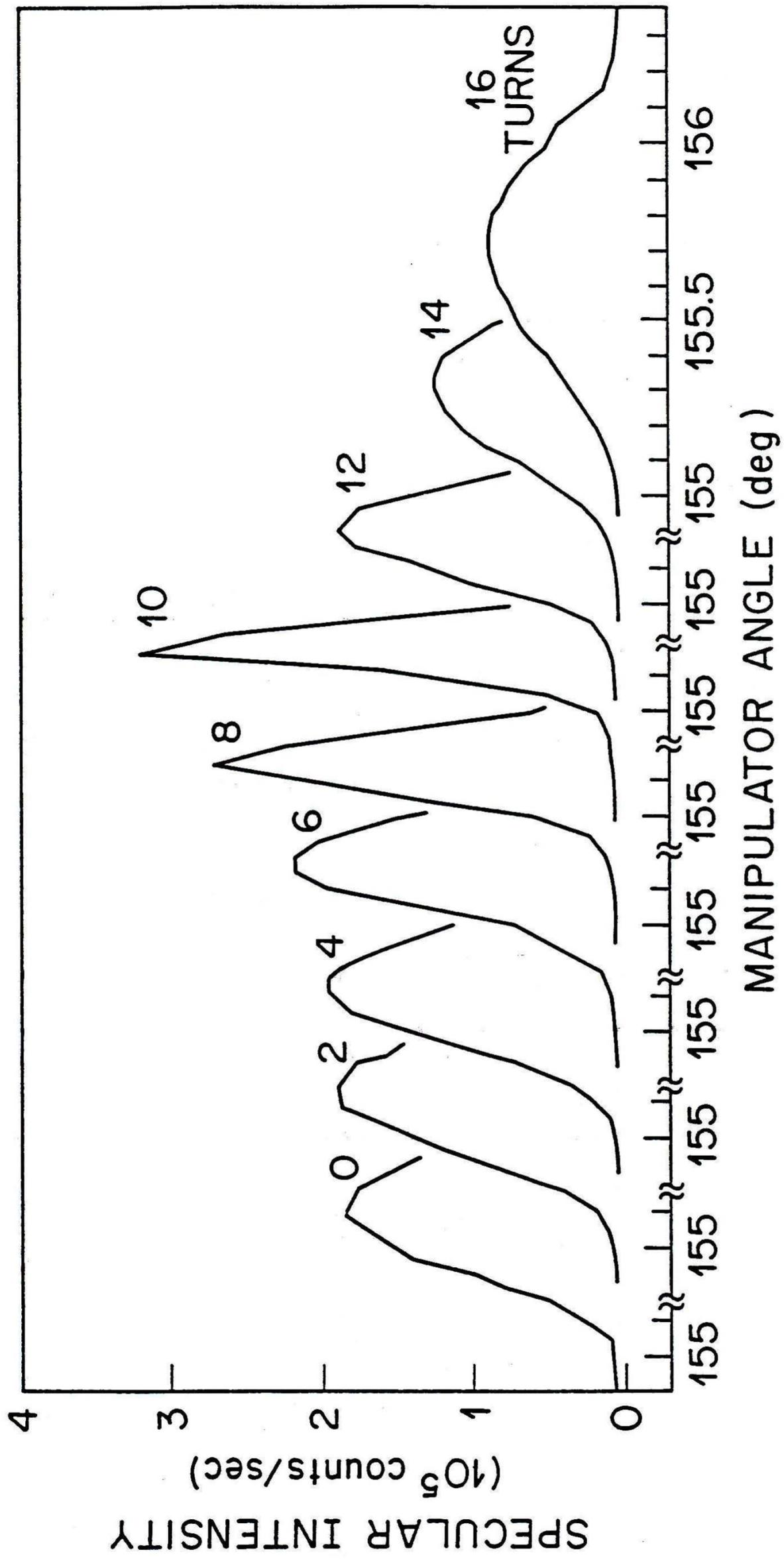

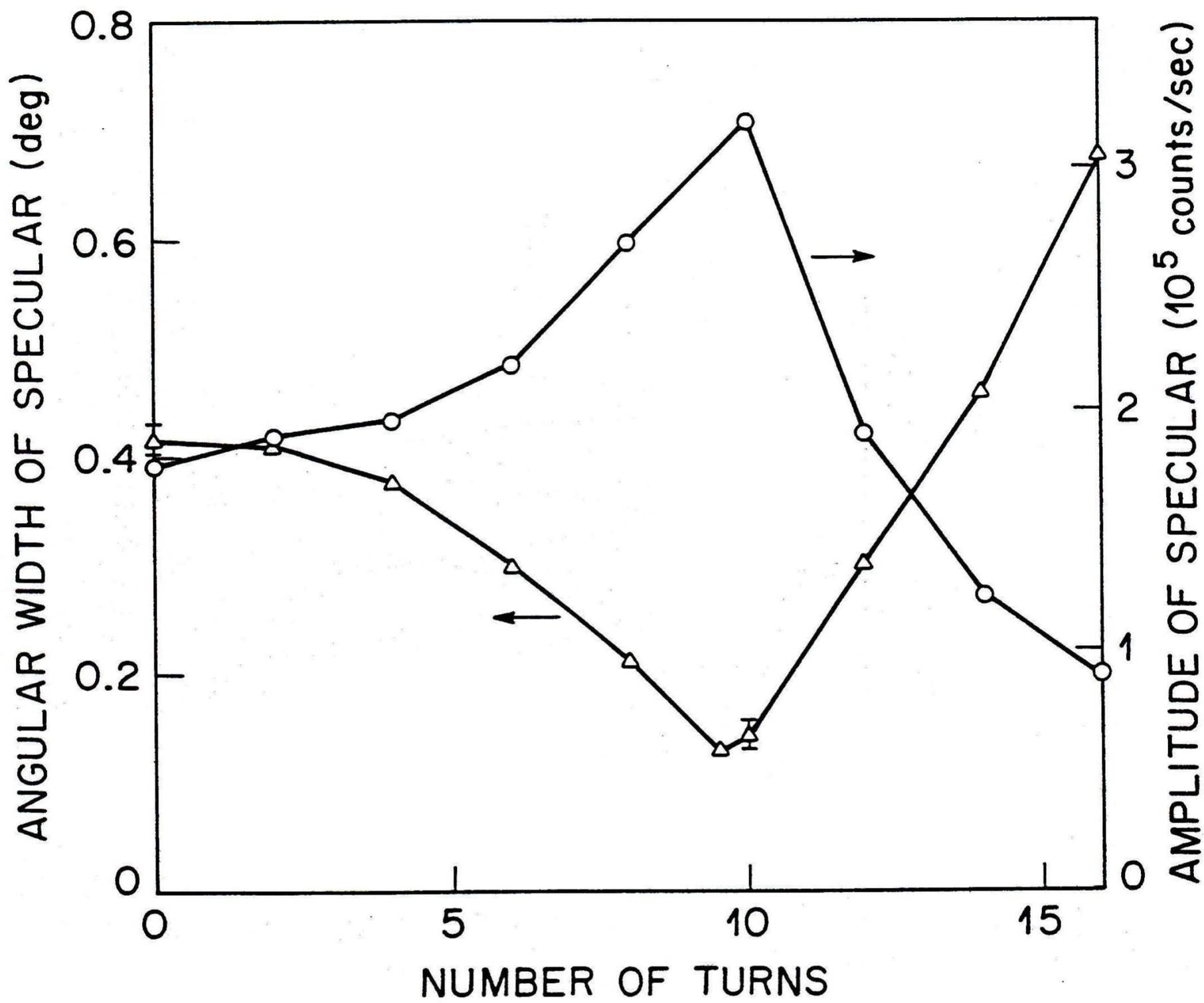

NUMBER OF TURNS

DJAK (8)

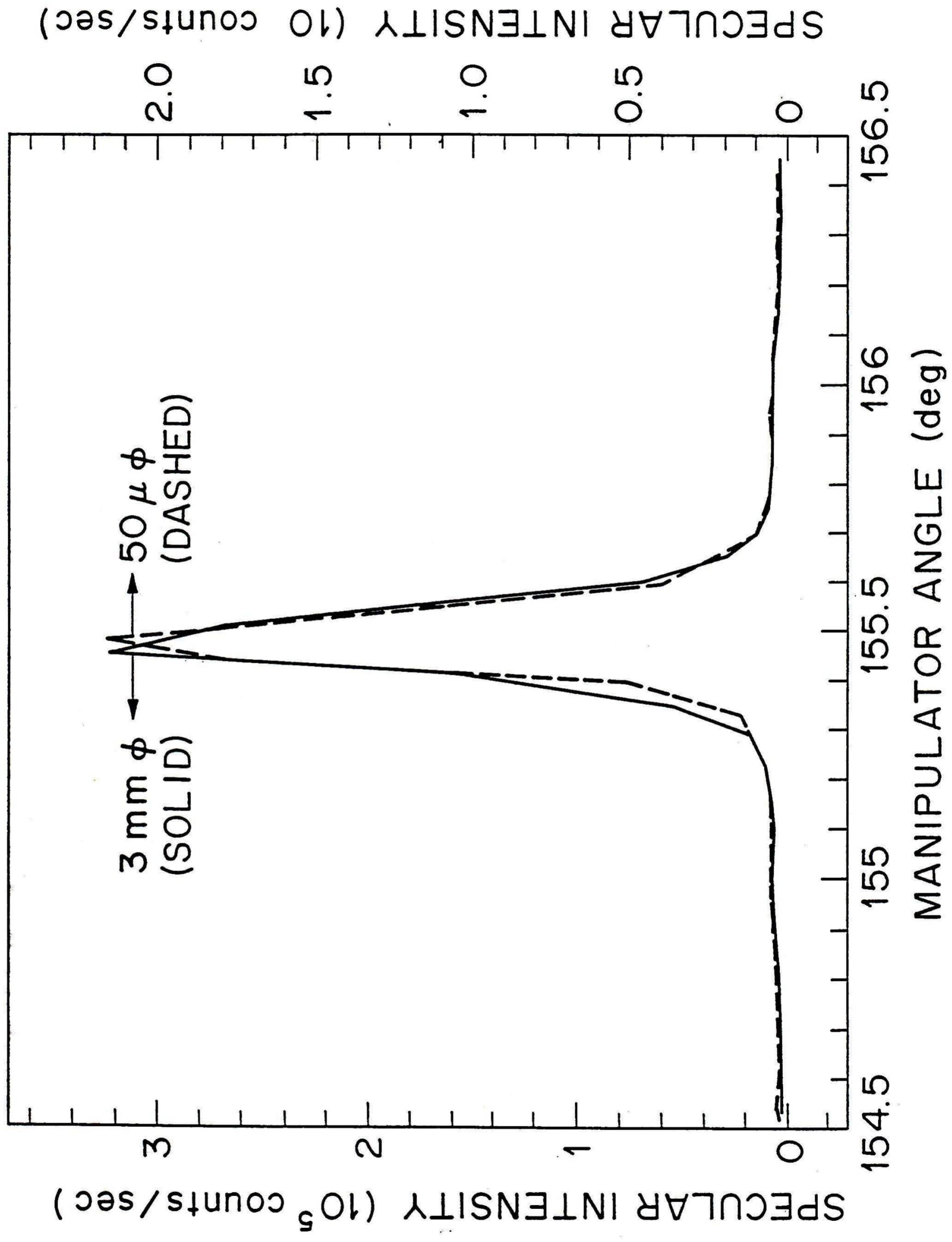

SPECULAR INTENSITY (10 counts/sec)

2.0   1.5   1.0   0.5   0

156.5   156   155.5   155   154.5

MANIPULATOR ANGLE (deg)

50 μ φ (DASHED)

3 mm φ (SOLID)

SPECULAR INTENSITY ($10^5$ counts/sec)

3   2   1   0

DO+K 9

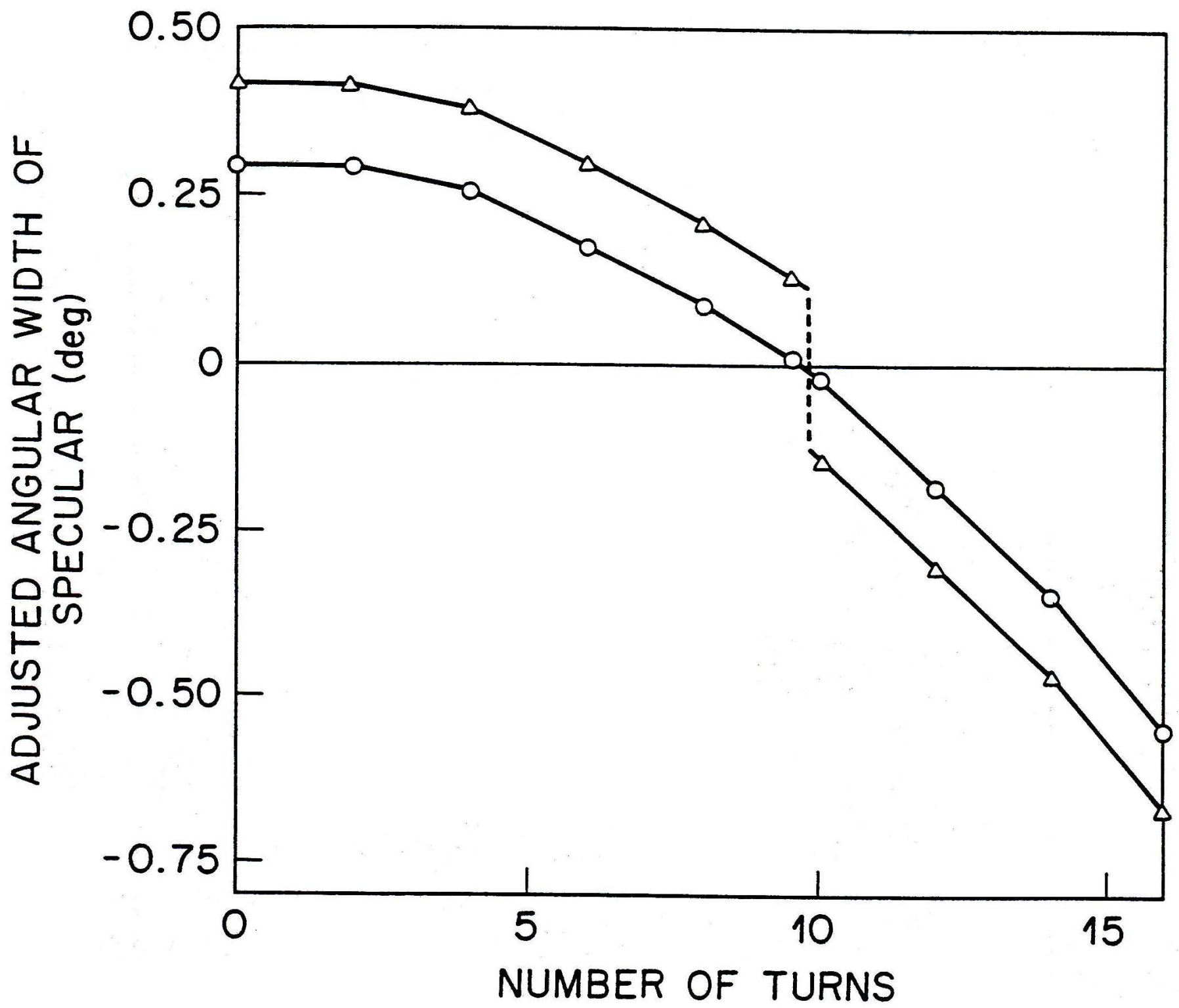

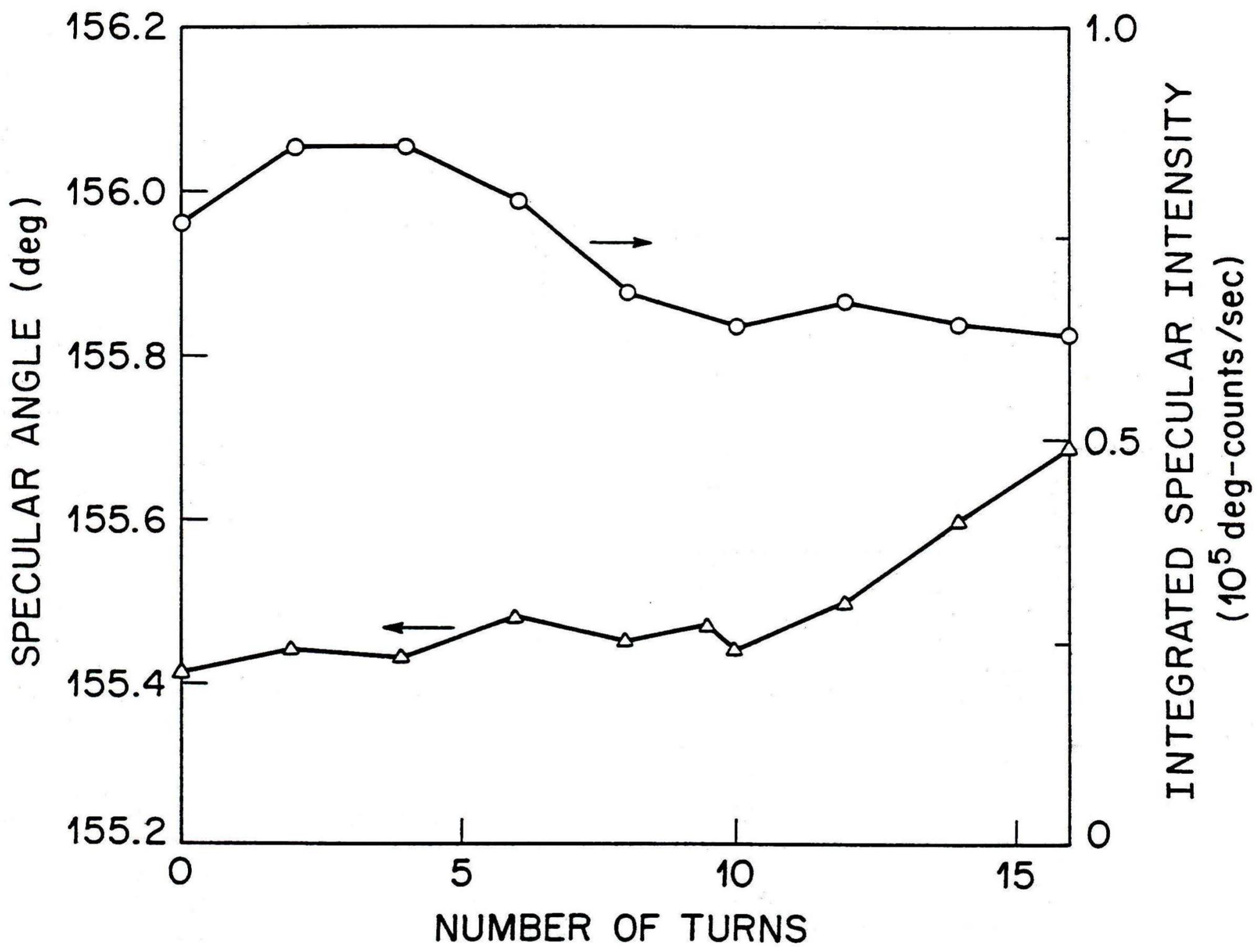

DOAK 11

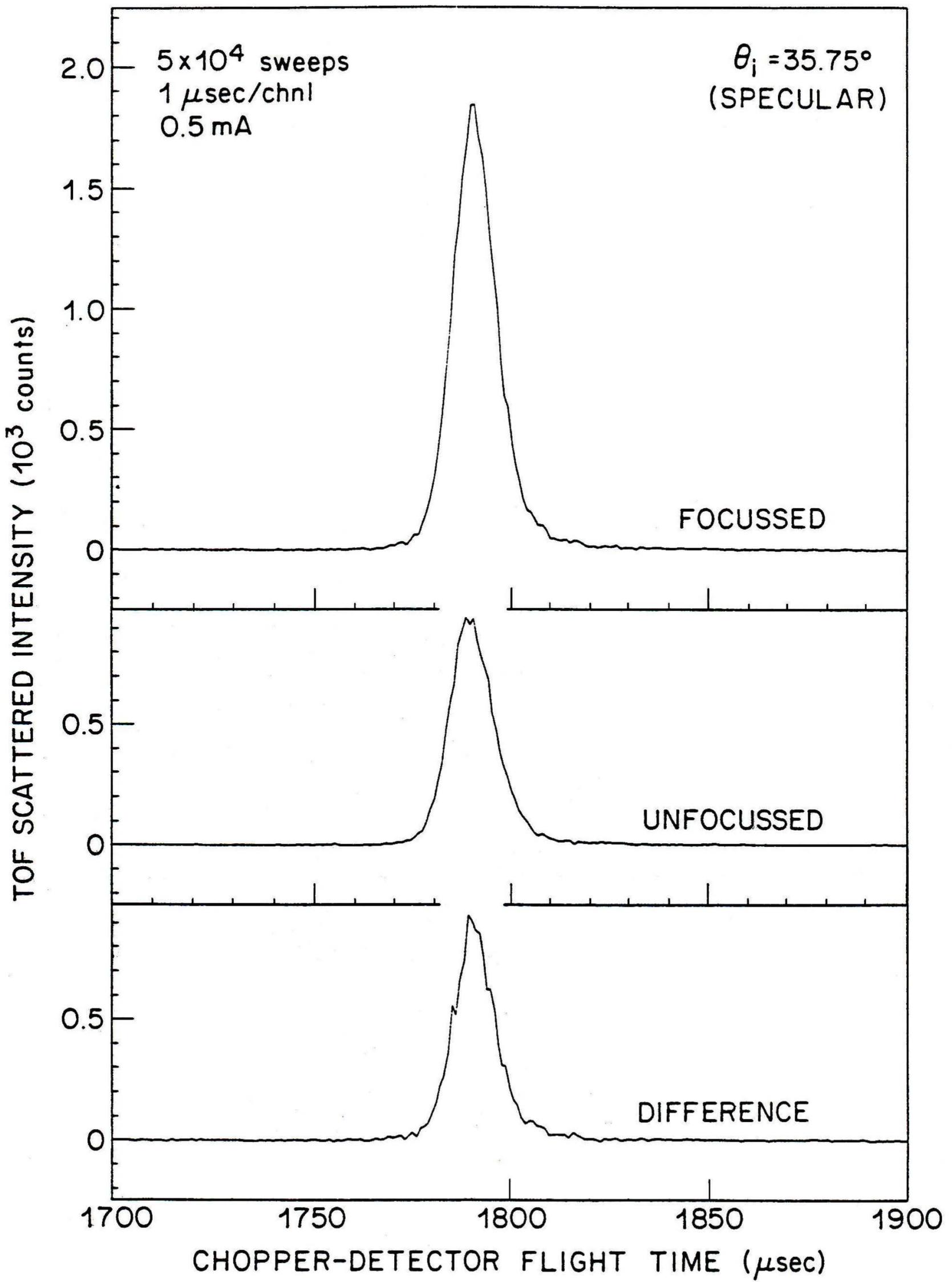

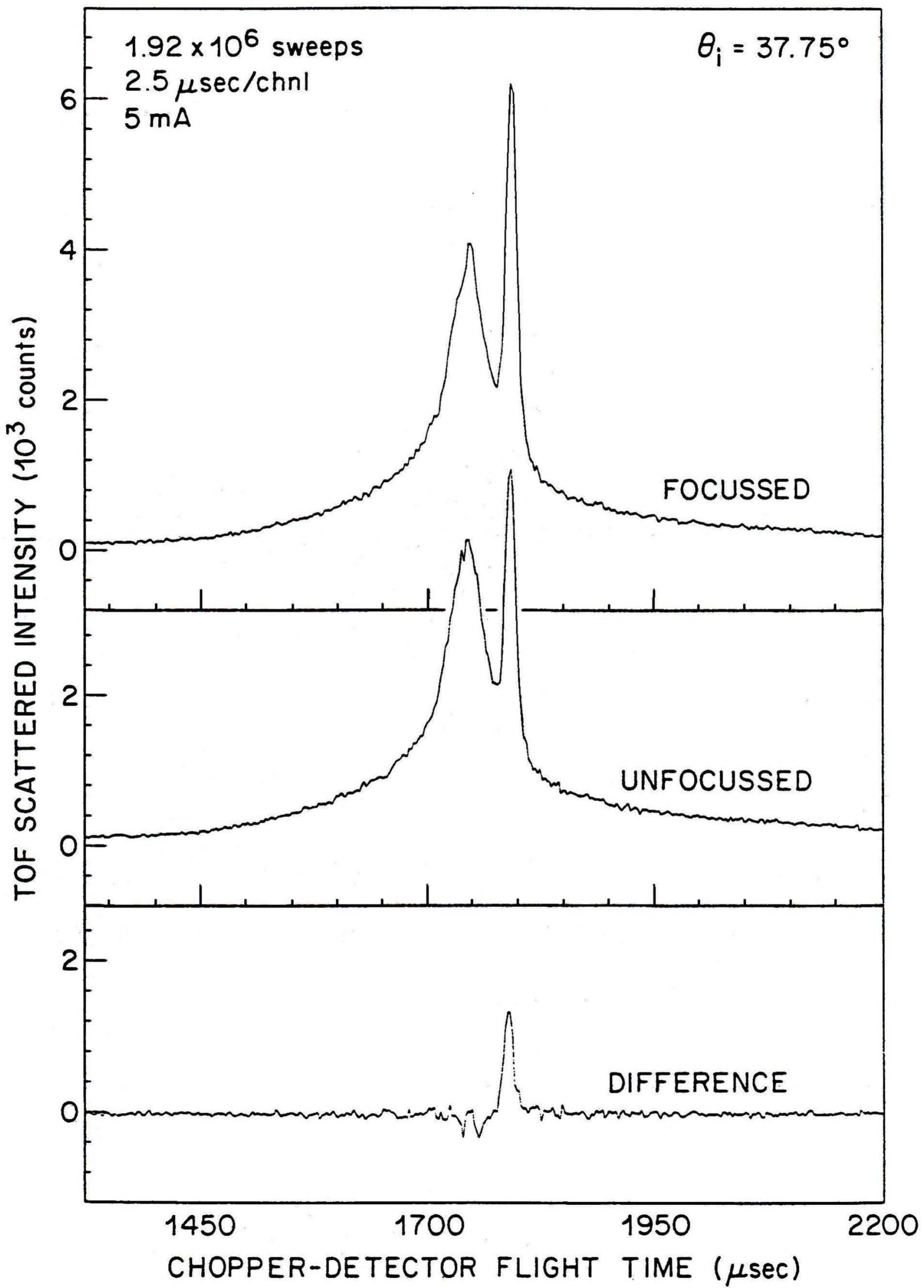

DOAK (13)

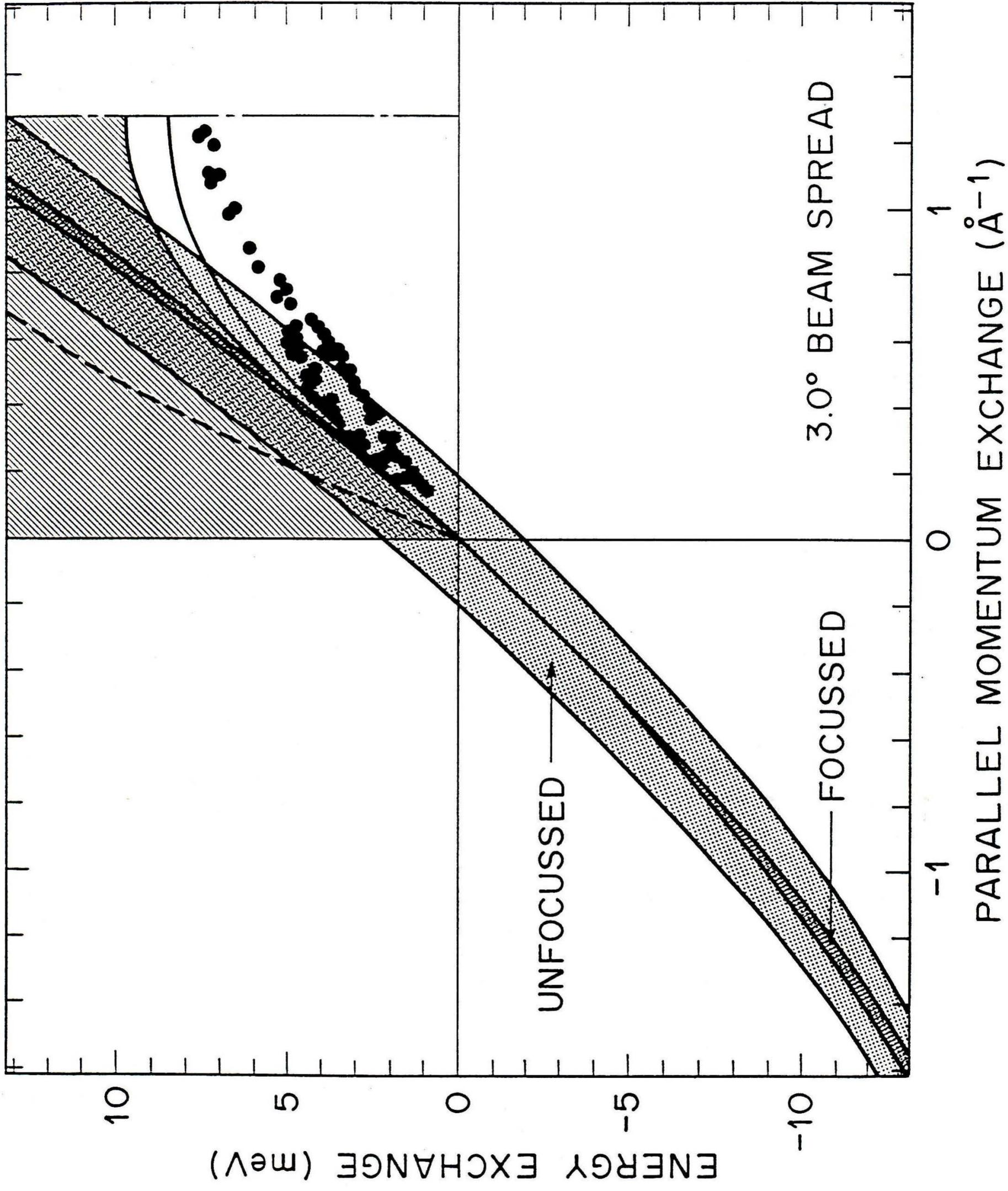

ENERGY EXCHANGE (meV)

PARALLEL MOMENTUM EXCHANGE (Å⁻¹)

3.0° BEAM SPREAD

UNFOCUSSED

FOCUSSED

DOAK 14

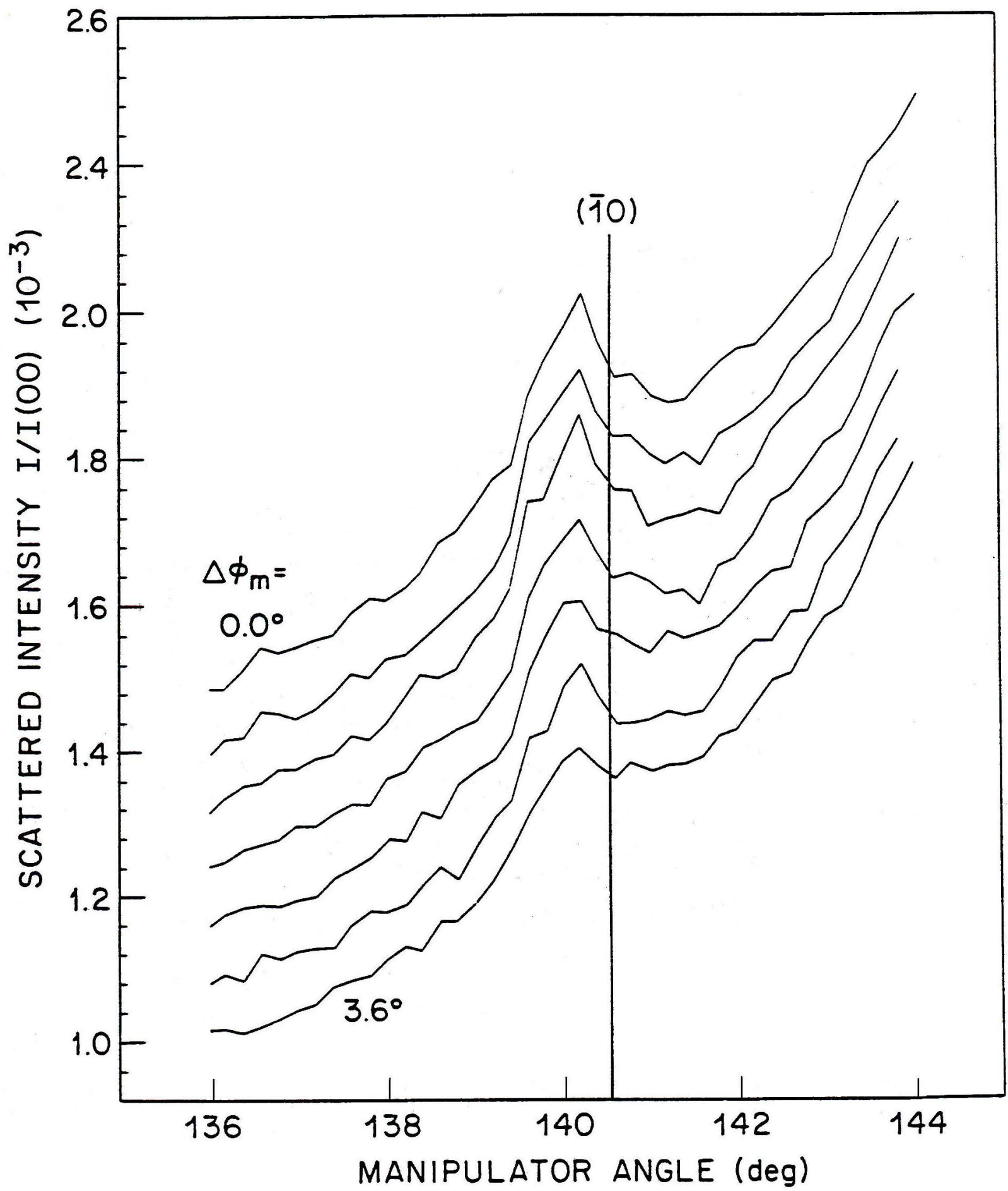

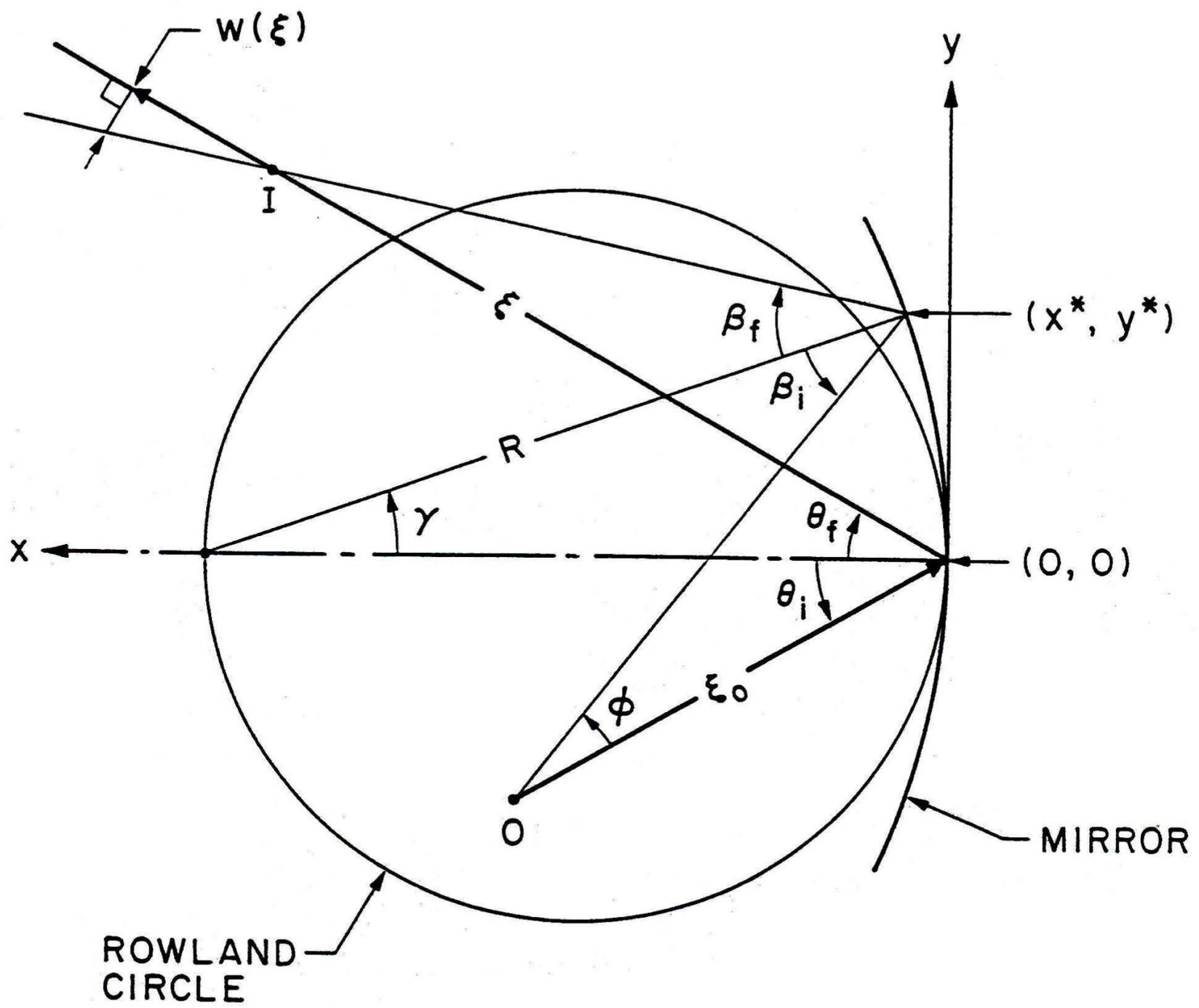

MIRROR

ROWLAND
CIRCLE

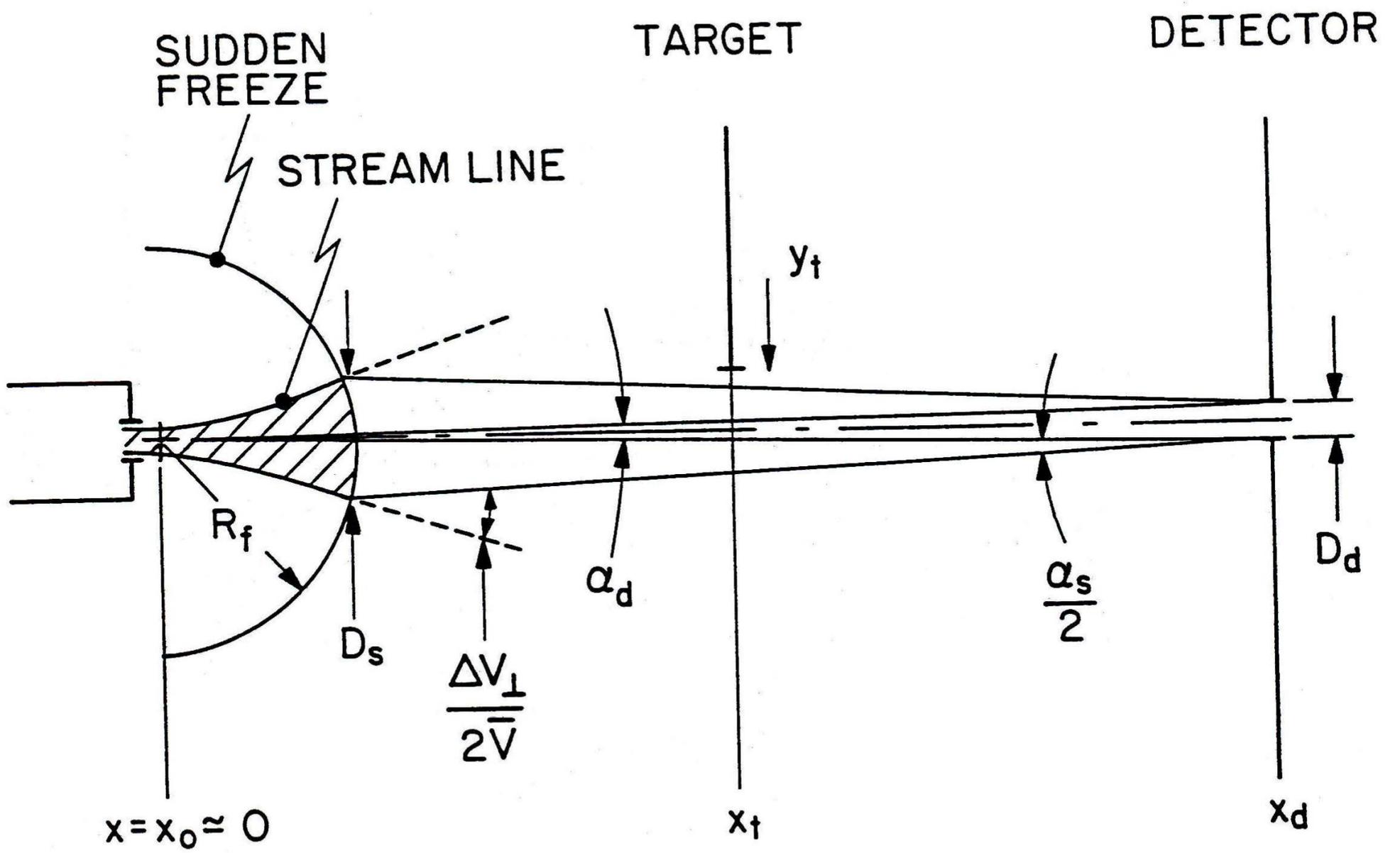

SUDDEN
FREEZE

STREAM LINE

TARGET

DETECTOR

$y_t$

$R_f$

$D_s$

$\dfrac{\Delta V_\perp}{2\overline{V}}$

$\alpha_d$

$\dfrac{\alpha_s}{2}$

$D_d$

$x = x_o \simeq 0$

$x_t$

$x_d$

DOAK 17

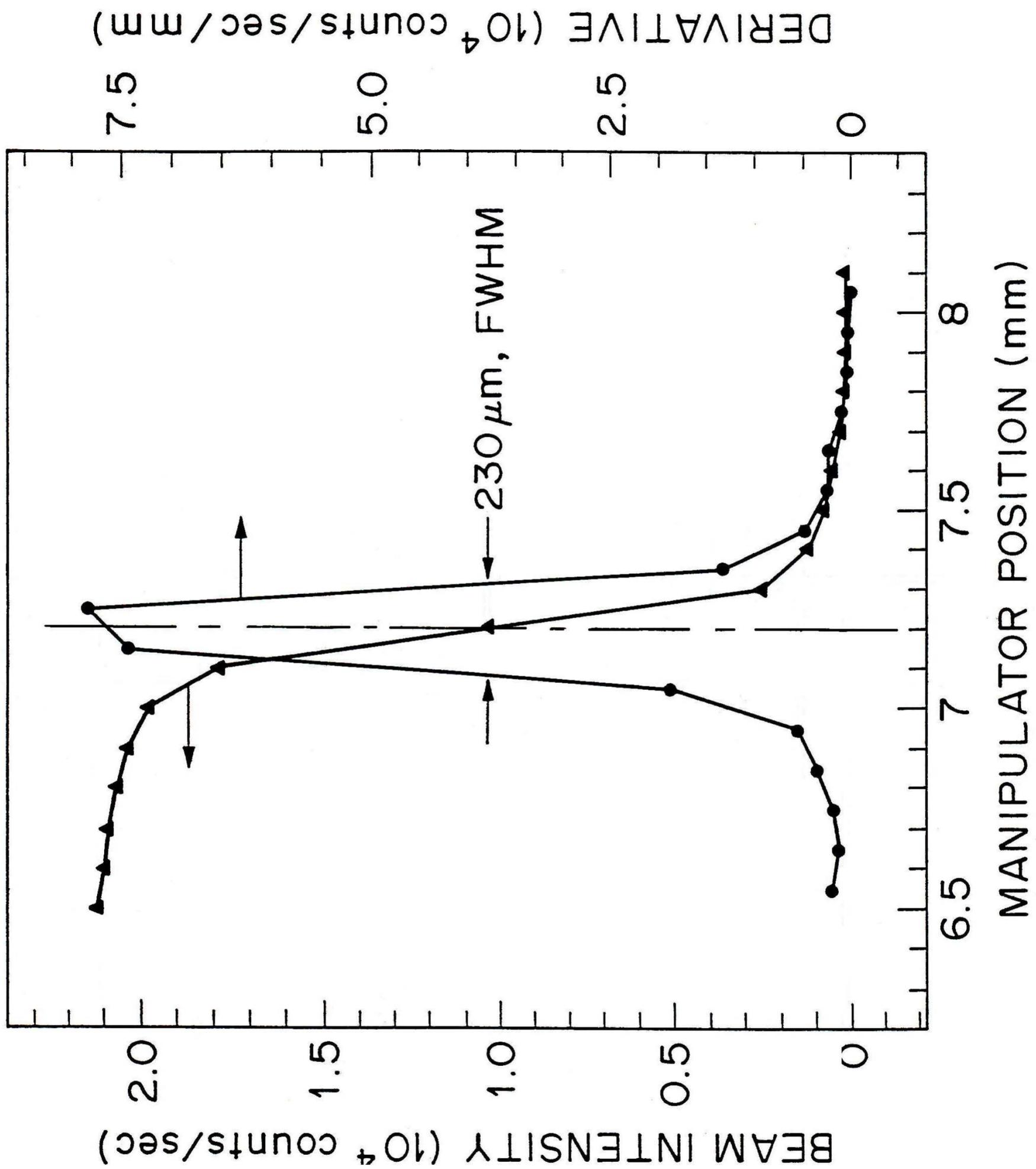